\documentclass[11pt,a4paper]{article}
\usepackage{jheppub}
\usepackage{amsthm,euscript,array,cancel}
\usepackage{slashed}

\title{Komar charge of ${\cal N}=2$ supergravity and its superspace
  generalization}

\author[a,b]{Igor Bandos,}

\author[c,d]{Patrick Meessen,}

\author[e]{Tom\'as Ort\'{\i}n}

\affiliation[a]{Department ofTheoretical
  Physics, University of the Basque Country UPV/EHU,\\ P.O. Box 644, 48080
  Bilbao, Spain}

\affiliation[b]{IKERBASQUE,\\ Basque Foundation for Science,
  48011, Bilbao, Spain}

\affiliation[c]{Departament of Physics, Universidad de Oviedo,\\
  C/ Leopoldo Calvo Sotelo 18, E-33007 Oviedo, Spain}

\affiliation[d]{Instituto Universitario de Ciencias y Tecnolog\'{\i}as
  Espaciales de Asturias (ICTEA),\\
  C/ de la Independencia 13, E-33004 Oviedo, Spain}

\affiliation[e]{Instituto de F\'{\i}sica Te\'orica UAM/CSIC,\\
  C/ Nicol\'as Cabrera, 13--15, C.U.~Cantoblanco, E-28049 Madrid, Spain}

\emailAdd{Igor.Bandos@ehu.eus}

\emailAdd{meessenpatrick@uniovi.es}

\emailAdd{Tomas.Ortin@csic.es}

\abstract{We find the defining equations for a Killing vector and its
  superpartner (called the generalized Killing spinor) and use them to
  construct the generalized Komar superform of minimal ${\cal N}=2$ $d=4$
  supergravity using the superspace formulation. The superspace procedure
  presented here can be used for the construction of generalized Komar forms
  in more general supergravity theories. We also present the (more cumbersome)
  calculation of the generalized Komar 2-form and of the on-shell closed
  2-form used to prove the first law in the component formalism, as an
  independent confirmation of our main result.}


\begin{document}
\maketitle

%%%%%%%%%%%%%%%%%%%%%%%%%%%%%%%%%%%%%%%%%%%%%%%%%%%%%%%%%%%%%%%%%%%%%%
%%%%%%%%%%%%%%%%%%%%%%%%%%%%%%%%%%%%%%%%%%%%%%%%%%%%%%%%%%%%%%%%%%%%%%
%%%%%%%%%%%%%%%%%%%%%%%%%%%%%%%%%%%%%%%%%%%%%%%%%%%%%%%%%%%%%%%%%%%%%%
%%%%%%%%%%%%%%%%%%%%%%%%%%%%%%%%%%%%%%%%%%%%%%%%%%%%%%%%%%%%%%%%%%%%%%
\section{Introduction}
%%%%%%%%%%%%%%%%%%%%%%%%%%%%%%%%%%%%%%%%%%%%%%%%%%%%%%%%%%%%%%%%%%%%%%
%%%%%%%%%%%%%%%%%%%%%%%%%%%%%%%%%%%%%%%%%%%%%%%%%%%%%%%%%%%%%%%%%%%%%%
%%%%%%%%%%%%%%%%%%%%%%%%%%%%%%%%%%%%%%%%%%%%%%%%%%%%%%%%%%%%%%%%%%%%%%
%%%%%%%%%%%%%%%%%%%%%%%%%%%%%%%%%%%%%%%%%%%%%%%%%%%%%%%%%%%%%%%%%%%%%%

The Komar form \cite{Komar:1958wp}, a 2-form based on a Killing vector which
is closed upon use of the vacuum Einstein equations, is commonly used to
define/compute the gravitational conserved charges associated to Killing
vectors in spacetimes that admit them by integrating it over the boundary at
spatial infinity (a 2-sphere in 4-dimensional asymptotically-flat spacetimes):
the so-called Komar integral. The closedness of the Komar form implies that
one could perform these integrals over any other surface topologically
equivalent to the 2-sphere at infinity.\footnote{This behaviour (a Gauss law)
  may seem to suggest that the gravitational charges (such as the mass) are
  localized, which is not true. The fact that the Komar charge satisfies a
  Gauss law is a special property of spacetimes that have Killing vectors. In
  the general case, only asymptotic Killing vectors exist and one must
  use other methods \cite{Arnowitt:1962hi,Abbott:1981ff}.}

The Komar form can still be used in such a way in the presence of matter, but,
in general, it is not closed anymore and the integral must be performed at
infinity.

The closedness of the Komar form may be recovered by adding certain terms to
it to define a new, on-shell closed 2-form which we will call
\textit{generalized Komar form}.  This is totally unnecessary if our only goal
is to compute conserved charges and the integral at infinity can be performed,
but the Komar form and its generalizations have, at least, another important
use that depends on its closedness:\footnote{Other uses are discussed, for
  instance, in \cite{Ballesteros:2024prz}.} the derivation of Smarr formulae
in black-hole spacetimes \cite{Smarr:1972kt}, an idea that goes back to the
seminal papers by Bardeen, Carter and Hawking
\cite{Bardeen:1973gs,Carter:1973rla}. Smarr formulae can be derived through
scaling arguments under some assumptions, but there is much to learn from
direct, rigorous proofs. This makes the construction of the generalized Komar
forms of different theories an important task.

In the pioneering paper \cite{Simon:1984qb}, Simon investigated sufficient
conditions for matter coupled to gravity to admit a generalized Komar 2-form
and found, amongst other things, that minimal $\mathcal{N}=2$, $d=4$
supergravity (including the fermions) satisfied his criteria and constructed
its generalized Komar form, albeit in a particular gauge.
\par
As the existence of a generalized Komar form is ultimately an on-shell
property, and one can relate (classes of) solutions by using the symmetries of
the equations of motion, one would venture that it should be possible to find
a generalized Komar form that is invariant under all the symmetries of the
equations of motion. In a supergravity theory, this not only means that one
should be able to find a formulation of the generalized Komar form that is
duality invariant, but also invariant under supersymmetry. Even though Simon
\cite{Simon:1984qb} mentions the possibility of giving a purely magnetic
formulation of his generalized Komar 2-form, his criterion for the invariance
of the fields under the action of the Killing vector by means of the Lie
derivative is up to compensating local Lorentz transformations (i.e.~he uses
the so-called Kosmann or Lie-Lorentz derivative
\cite{Kosmann:1971,Vandyck:1988ei,Ortin:2002qb}) (see also
\cite{Fatibene2011}), but ignores other possibilities, such as compensating
gauge \cite{Prabhu:2015vua,Elgood:2020svt} and/or supersymmetry
transformations \cite{Vandyck:1988gc}.  Once opened up to the idea of a
generalized Komar form in supergravity being invariant under supersymmetry,
however, the question is how to construct it in the most general way possible.
\par
In \cite{Bandos:2023zbs}, two of the present authors constructed the
supersymmetric generalized Komar 2-form for $\mathcal{N}=1$, $d=4$
supergravity, using the Lie-covariant derivative formalism
\cite{Elgood:2020svt,Ortin:2022uxa,Ortin:2024mmg} and the notion of a Killing
supervector in on-shell supergravity superspace
\cite{Buchbinder:1995uq}.\footnote{ See also
  \cite{Vandyck:1989ai,Kuzenko:2012vd,Kuzenko:2015lca,Howe:2015bdd,Howe:2018lwu,Kuzenko:2019tys,Chandia:2022uyy}
  as well as the slightly different approach of
  \cite{Figueroa-OFarrill:2007omz}.  } Using the Killing supervector, it is
paramount that the Killing vector has a superpartner, called the generalized
Killing spinor, which contributes to the supergravity generalization of the
usual Killing equation, by adding terms that are bilinears in the gravitino
fields and the generalized Killing spinor
\cite{Vandyck:1988gc,Buchbinder:1995uq}. The generalized Killing and Killing
spinor equations, as well as the supersymmetry properties of the Killing
vector and spinor, follow from the defining superfield equations for the
Killing supervector in $\mathcal{N}=1$, $d=4$ superspace, taking into account
the superspace constraints for minimal supergravity. The sought for generalized
Komar 2-form, i.e.~the on-shell closed and supersymmetric 2-form, was then
seen to be equal to the Noether-Wald 2-form for the diffeomorphism
\cite{Wald:1993nt,Jacobson:2015uqa} associated to the Killing
vector.
\par
In \cite{Bandos:2024pns} it was further observed that the $\mathcal{N}=1$
$d=4$ generalized Komar 2-form corresponds to the lowest component of a {\em
  super-Komar form}, by which we mean a closed, bosonic 2-superform in the
on-shell minimal $\mathcal{N}=1$, $d=4$ supergravity superspace: as
supersymmetry acts via the super-Lie derivative on the superform, its
closedness guarantees that the super-Komar form, whence also the generalized
Komar 2-form, is in fact invariant under supersymmetry (up to a total
derivative).

{\bf Actually}
%The results of
\cite{Bandos:2023zbs} have provided the proof-of-concept-type
results on the existence, conservation and supersymmetry invariance of the
generalized Komar forms associated to Killing supervectors of supergravity.  Indeed, one
of the possible applications of such kind of results is the study of
supersymmetric black-hole (BH) thermodynamics for black holes with fermionic
hair. However, no-go results for the existence of such BH solutions in
${\cal N}=1$ supergravity were proven in \cite{Gueven:1980be} (after these had
been conjectured in \cite{Cordero:1978ud}).  Supersymmetric BH solutions with
non-trivial fermionic hair do, however, exist in ${\cal N}=2$ supergravity
\cite{Gueven:1982tk,Aichelburg:1983ux}, so that the simplest supersymmetric BH
thermodynamics that can be developed on the basis of the approach of
\cite{Bandos:2023zbs} is the one corresponding to ${\cal N}=2$
supergravity. In this article we will construct the supersymmetric invariant
generalized Komar 2-form for minimal $\mathcal{N}=2$, $d=4$ supergravity, leaving possible
applications for future work.

Seeing that supersymmetry is the underlying building block of the Killing
supervector and the (super-)Komar form, it seems natural to use superspace
techniques in order to generalize the $\mathcal{N}=1$ approach of
\cite{Bandos:2023zbs} to the ${\cal N}=2$ case.  This generalization turns out
to be not straightforward, however, a fact that was, based on prior experience
with the construction of generalized Komar forms in purely bosonic theories,
to be expected: the usual Noether-Wald form for a diffeomorphism
\cite{Wald:1993nt} for gravity coupled to matter, is in general not an
on-shell closed codimension-2 form. It is, however, known how to create a
generalized Komar form by adding appropriate terms to the Noether-Wald form
\cite{Liberati:2015xcp,Ortin:2021ade,Mitsios:2021zrn,Ortin:2024mmg}).

In the Lie-covariant approach
\cite{Elgood:2020svt,Ortin:2022uxa,Ortin:2024mmg}, the fact that a given field
is invariant under the action of a Killing vector up to possible compensating
gauge transformations, is codified by the existence of momentum maps
associated to {\bf this} Killing vector, one for each possible gauge
transformation. For example, from the point of view of the action of minimal
$\mathcal{N}=1$, $d=4$ supergravity, there are two momentum maps: one
associated to local Lorentz transformations and another one to the
supersymmetry transformations, which turns out to correspond to the
generalized Killing spinor. In $\mathcal{N}=2$, the addition of the Maxwell
field with its gauge invariance, then naturally leads to the introduction of
the so-called electric momentum map. Combining superspace techniques with the
ideas of the Lie-covariant approach, leads naturally to the use of superfield
momentum maps associated to the basic superfield gauge symmetries in the
on-shell $\mathcal{N}=2$, $d=4$ superspace with the minimal constraints. The
straightforward application of the construction of \cite{Bandos:2023zbs} then
gives a superform that is not on-shell closed.

The solution to the aforementioned non-closedness goes hand in hand with the
solution of another problem, namely the lack of duality invariance: as the
action is not invariant under electromagnetic duality, the possibility of
having a (super-)Komar form that is duality invariant seems to be excluded.
In \cite{Ortin:2022uxa}, however, it was shown that the introduction of a
magnetic momentum map completes the generalized Komar form in a
duality-invariant form.\footnote{Although this is not manifest in
  \cite{Ortin:2022uxa}, the magnetic momentum map is secretly related to the
  gauge invariance of the dual (magnetic) vector field that one can define
  on-shell. The presence of a magnetic momentum map is very natural in the
  context of fully democratic formulations (see, for instance,
  \cite{Meessen:2022hcg}) and PST formulation \cite{Pasti:1996vs,DallAgata:1997yxl,DallAgata:1998ahf}. In our case, we would need a fully democratic
  formulation of minimal $\mathcal{N}=2$, $d=4$ supergravity
  \cite{deWit:2005ub}.} The introduction of a superfield magnetic momentum map
in the on-shell $\mathcal{N}=2$, $d=4$ superspace with the minimal
constraints, then allows for the construction of the super-Komar form in a
very convenient manner, one that should be straightforwardly generalizable to
other, more complex supergravities.
\par
The outline of this article is as follows: in sections 2 and 3, we will detail
how to obtain the on-shell superspace description of minimal $\mathcal{N}=2$
$d=4$ supergravity, from a first-order action. In section 4, we will discuss
the definitions of the Killing supervector in on-shell superspace, introduce
the superspace momentum maps and give the on-shell closed super-Komar form and
its lowest component, the generalized Komar 2-form. In section 5, we will
briefly compare the superspace construction with the construction of the
generalized Komar 2-form based on the various Noether currents/charges, and
show that the superspace construction is the more efficient method.  In
appendix \ref{sec:Appendix}, we will outline the construction of the
generalized Komar 2-form using the spacetime component formalism along the
lines of \cite{Elgood:2020svt,Mitsios:2021zrn}, which is far more involved
than the superspace approach; for the moment the only advantage of the
spacetime component approach is that it allows for the construction of an
on-shell closed 2-form along the lines of
\cite{Wald:1993nt,Prabhu:2015vua,Jacobson:2015uqa,Elgood:2020svt,Ortin:2022uxa},
that gives rise to the first law of Killing horizon thermodynamics; the
corresponding (first law) 2-form for minimal $\mathcal{N}=2$, $d=4$ is
constructed in section \ref{appsec:FirstLaw}.
\par
Finally, in section 6 we present our conclusions and outlook.

%%%%%%%%%%%%%%%%%%%%%%%%%%%%%%%%%%%%%%%%%%%%%%%%%%%%%%%%%%%%%%%%%%%%%%
%%%%%%%%%%%%%%%%%%%%%%%%%%%%%%%%%%%%%%%%%%%%%%%%%%%%%%%%%%%%%%%%%%%%%%
%%%%%%%%%%%%%%%%%%%%%%%%%%%%%%%%%%%%%%%%%%%%%%%%%%%%%%%%%%%%%%%%%%%%%%
%%%%%%%%%%%%%%%%%%%%%%%%%%%%%%%%%%%%%%%%%%%%%%%%%%%%%%%%%%%%%%%%%%%%%%
\section{Minimal ${\cal N}=2$ sugra: from the first-order action to the generalized action}
%%%%%%%%%%%%%%%%%%%%%%%%%%%%%%%%%%%%%%%%%%%%%%%%%%%%%%%%%%%%%%%%%%%%%%
%%%%%%%%%%%%%%%%%%%%%%%%%%%%%%%%%%%%%%%%%%%%%%%%%%%%%%%%%%%%%%%%%%%%%%
%%%%%%%%%%%%%%%%%%%%%%%%%%%%%%%%%%%%%%%%%%%%%%%%%%%%%%%%%%%%%%%%%%%%%%
%%%%%%%%%%%%%%%%%%%%%%%%%%%%%%%%%%%%%%%%%%%%%%%%%%%%%%%%%%%%%%%%%%%%%%

Besides the graviton described by the vierbein 1-form
$e^a=dx^{\mu} e_{\mu}{}^{a}(x)$ and the two gravitini described by the
fermionic spinor 1-form\footnote{ In this paper we will use, following
  \cite{Ortin:2015hya}, the Majorana spinor notation with hidden $SU(2)$
  indices.  The Weyl spinor notation with explicit $\alpha=1,2$,
  $\dot{\alpha}=1,2$ and explicit SU(2) indices $i=1,2$ will be used in a few
  equations with the aim to clarify the Majorana spinor expressions and to
  reflect the relation of the notation of \cite{Ortin:2015hya} with the ones
  of \cite{Bandos:2019lps} and references therein.  }
$\psi= \sqrt{2}(\psi_{\alpha}^i, \bar{\psi}^{\dot{\alpha}}_i)=dx^\mu
\psi_\mu$, the minimal ${\cal N}=2$ supergravity multiplet contains a vector
field $A=dx^\mu A_\mu=e^a\ A_{a}(x)$.
\par
In order to write down a first-order action, the above fields must be
complemented with an independent spin connection 1-form
$\omega^{ab}=dx^\mu \omega_\mu^{ab}$ and an independent anti-symmetric tensor
field $F_{ab}(x)$.  The spin connection is used to define the Riemann
curvature 2-form and the covariant exterior derivative of the gravitino 1-form
as\footnote{ In the main part of this article we will use the convention that
  the (covariant) exterior derivative and the interior product act from the
  right, as this is much more convenient for superspace calculation, that are
  used extensively in this article. This convention for instance means that
  $d\psi = dx^{\nu}\wedge d\psi_{\nu}= dx^{\nu}\wedge dx^{\mu}\
  \partial_{\mu}\psi_{\nu}$.  }
\begin{eqnarray}
  \label{Rab=N2}
  R^{ab}
  & =&
  d\omega^{ab} -\omega^{ac}\wedge \omega_c{}^{b}
  \; =\; \tfrac{1}{2}\ e^d\wedge e^c\ R_{cd}{}^{ab}
  \; , \\
  \label{cDpsi=N2}
  {\cal D}\psi
  & =&
  d\psi -\tfrac{1}{4}\ \omega^{ab}\wedge \gamma_{ab}\psi
  \; =\;
  d\psi -\tfrac{1}{4}\ \slashed{\omega}\wedge \psi
  \; ,
  %d\psi -\frac 1 4 \omega\!\!\!/{}\;\wedge \psi
\end{eqnarray}
and $F_{ab}(x)$ is used to construct the 2-form
$F_2=\tfrac{1}{2}\ e^b\wedge e^a\ F_{ab}$ as well as its Hodge dual
$*F_2=\tfrac{1}{4} e^b\wedge e^a \epsilon_{abcd} F^{cd}(x)$.
\par
Given the above ingredients, the first-order action reads
\begin{eqnarray}\label{S=SG=N2=}
 S^{{\cal N}=2}= \int_{M^4} {\cal L}_4^{{\cal N}=2}= \int\left(-\frac 1 2 \epsilon_{abcd}R^{ab}\wedge e^c\wedge e^d +
 2\bar{\psi}\wedge \gamma_5\gamma\wedge {\cal D}{\psi}- \right.
 \nonumber \\
 {} \nonumber \\
 \left. \hspace{2.2cm}- \frac 1 2 F_2\wedge *F_2 + (dA-J_m)\wedge (*F_2-\tilde{J}_e)  -\frac 1 2 J_m\wedge \tilde{J}_e\right)
\end{eqnarray}
where we furthermore used the matrix-valued 1-form
$\gamma\ :=\ e^{a}\gamma_{a}$ and the following notation for the
$SO(1,3)\times SU(2)$ invariant fermionic bilinears\footnote{The bilinears
  used in this paper, $J_m,\tilde{J}_{e}$ are related to those defined in
  \cite{Ortin:2015hya}, $\mathcal{J}_{(e)},\mathcal{J}_{(m)}$ by
  $J_{m}=\mathcal{J}_{(e)}$ and $\tilde{J}_{e}=\star \mathcal{J}_{(m)}$. The
  reason for this change is $dJ_{e}$ occurs, in a natural way, as the electric
  current source and $dJ_{m}$ as a dual magnetic current source in the Maxwell
  equations and Bianchi identities.}
\begin{equation}
  \label{Jm=}
  J_m= i\bar{\psi}\wedge \tau^2 \psi \; ,  \qquad  \tilde{J}_e  = -(\bar{\psi}\wedge \gamma_5\tau^2\wedge \psi)\; ,
\end{equation}

\noindent
where

\begin{eqnarray}  \label{atu2}
\tau^2=\left(\begin{matrix} i\epsilon_{ij} & 0 \cr
0 & - i\epsilon^{ij} \end{matrix}\right)\; , \qquad \epsilon^{ij} = \left(\begin{matrix} 0 & 1 \cr
-1 & 0 \end{matrix}\right)= -\epsilon_{ij} .
\end{eqnarray}

The equation of motion for the spin connection $\omega^{ab}=dx^\mu\omega_\mu^{ab}$ expresses the torsion 2-form in terms of the other gravitino bilinear,

\begin{equation}
  \label{Ta=N2x}
  T^a
  :=
  De^a
  :=
  de^a-e^b\wedge \omega_b{}^a
  = \frac i 2 \bar{{\psi}}\wedge \gamma^a{\psi}
  \; ,
\end{equation}
whereas the equation of motion for the antisymmetric tensor field $F_{ab}$
identifies it with a supersymmetric gauge field strength, namely
\begin{equation}
  \label{dA-Jm=F}
  F_{2}
  \; =\;
  dA \ -\ J_{m}
  \; .
\end{equation}
The equations \eqref{Ta=N2x} and \eqref{dA-Jm=F} can be substituted into
\eqref{S=SG=N2=}, resulting in a second order formalism action.

The variation of the action \eqref{S=SG=N2=} with respect to the 1-form gauge
potential $A=dx^\mu A_\mu(x)$ results in the generalized Maxwell equation

\begin{equation}
  \label{d(*F-Je)=0}
  d(*F_2-\tilde{J}_e)
  \; =\;
  0
  \; .
\end{equation}
% Of course this cannot be substituted back into the action if we wish to use
% this to obtain other equations.

\noindent
The other dynamical equations are the Einstein and the covariantized
Rarita-Schwinger equations which can be written in terms of differential forms
as
\begin{eqnarray}\label{EinEq=N2}
  {\mathbf E}_{a3}
  & =&
  \epsilon_{abcd}R^{bc}\wedge e^d+2\bar{\psi}\wedge \gamma_5\gamma_a {\cal D}{\psi}
  +
  \tfrac{1}{2} \left(F_2\wedge \imath_{a}*F_2 - \imath_{a}F_{2}\wedge *F_2  \right)
  \; =\;
  0
  \; , \\
  % & & \nonumber \\
  \label{RSeq=N2M}
  \mathbf{E}_{3}
  & =&
  4\gamma_5\ \gamma\wedge  {\cal D}{\psi}-2i   (*F_2+i\gamma_5F_2) \wedge \tau^2{\psi}
  \; =\;
  0\; .
\end{eqnarray}
% {\bf{[***Differences in sign of the second term in \eqref{RSeq=N2M} with
% (2.16c) of the Tom\'{a}s file is due to $d$ and ${\cal D}$ acting from the
% right, see footnote \ref{d=left}.***] }}

The action and the equations of motion are invariant under the following
supersymmetry transformations of the fields of supergravity multiplet
\begin{equation}\label{N2=susy}
  \delta_\epsilon e^a
  \; =\;
  - i\bar{\epsilon}\gamma^{a}\psi \; ,\qquad
   \delta_\epsilon{\psi}
   \; =\; {\cal D}\epsilon+\frac{1}{8}\ F_{ab} \gamma^{ab}\ \gamma\tau^2\epsilon
   \; ,\qquad
   \delta_\epsilon A
   \; =\;  \imath_{\epsilon}J_m
   \; =\; -2i\bar{\epsilon}\tau^2\psi
   \, ,
 \end{equation}
%\begin{eqnarray}\label{N2=susy}
% \delta_\epsilon e^a=- i\bar{\epsilon}\gamma^a\psi^i \; , \qquad
% % \label{susy=psi=N2M}
%  \delta_\epsilon{\psi}={\cal D}\epsilon+\frac 1 8   F_{ab}  {\gamma}^{ab}\gamma^{(1)}\tau^2\epsilon\; ,
%  % \qquad
% %\\ \nonumber \\ \label{susy=A}
% \delta_\epsilon A =  i_\epsilon  J_m=  -2\bar{\epsilon}\tau^2\psi \, , \quad \\ \nonumber
%\end{eqnarray}
 supplemented with suitable supersymmetry transformations of the auxiliary
 fields $F_{ab}(x)$ and $\omega^{ab}$, which imply the following relations
 that are useful in the derivation of our results\footnote{ Throughout this
   article, we will use the notation $\doteq$ to denote identities that hold
   when the equations of motion are satisfied.  }
\begin{eqnarray}\label{omN2=susy}
  \tfrac{1}{2} \epsilon_{abcd} \delta_\epsilon\omega^{ab}\wedge  e^c\wedge e^d
  & \dot{=}&
  2\bar{\epsilon}(F_2+i\gamma_5*F_2)\tau^2\psi
  \, , \\
  % & & \nonumber \\
  \label{*F2=susy}
  \delta_\epsilon (*F_2)
  & \dot{=}&
  -2\bar{\psi}\wedge \gamma_5\tau^2\psi
  \; .
\end{eqnarray}

%\noindent

%%%%%%%%%%%%%%%%%%%%%%%%%%%%%%%%%%%%%%%%%%%%%%%%%%%%%%%%%%%%%%%%%%%%%%
%%%%%%%%%%%%%%%%%%%%%%%%%%%%%%%%%%%%%%%%%%%%%%%%%%%%%%%%%%%%%%%%%%%%%%
%%%%%%%%%%%%%%%%%%%%%%%%%%%%%%%%%%%%%%%%%%%%%%%%%%%%%%%%%%%%%%%%%%%%%%
%%%%%%%%%%%%%%%%%%%%%%%%%%%%%%%%%%%%%%%%%%%%%%%%%%%%%%%%%%%%%%%%%%%%%%
\subsection{Generalized action}
%%%%%%%%%%%%%%%%%%%%%%%%%%%%%%%%%%%%%%%%%%%%%%%%%%%%%%%%%%%%%%%%%%%%%%
%%%%%%%%%%%%%%%%%%%%%%%%%%%%%%%%%%%%%%%%%%%%%%%%%%%%%%%%%%%%%%%%%%%%%%
%%%%%%%%%%%%%%%%%%%%%%%%%%%%%%%%%%%%%%%%%%%%%%%%%%%%%%%%%%%%%%%%%%%%%%
%%%%%%%%%%%%%%%%%%%%%%%%%%%%%%%%%%%%%%%%%%%%%%%%%%%%%%%%%%%%%%%%%%%%%%

The advantage of the first-order action \eqref{S=SG=N2=} is that it can be
straightforwardly 'lifted' to the generalized action of the so-called {\it
  rheonomic} or {\it group manifold} approach to supergravity
\cite{Neeman:1978njh,Castellani:1991eu}, which in its turn can be used to
obtain the superspace constraints of supergravity
\cite{Neeman:1978njh,Castellani:1991eu} (and, actually, all their consistency
conditions).

The generalized action is obtained from the first-order action
\eqref{S=SG=N2=} by straightforward lifting its Lagrangian top form to
superspace by the transparent prescription of
$$x^\mu \mapsto Z^M=(x^\mu,\theta)=(x^\mu,\theta^{\check{\alpha}{\check i}},
\bar{\theta}{}^{\check{\dot{\alpha}}}_{{\check i}})$$ and
%\begin{eqnarray}\label{e->E}
%  e^a\mapsto E^a=dZ^ME_M^a(Z) , \qquad \psi \mapsto {\cal E} = \sqrt{2}(E_{\alpha}^{ i},\bar{E}^{\dot{\alpha}}_{i})
%  =dZ^M{\cal E} _M(Z) ,  \qquad \\    \nonumber \\\omega^{ab}(x)\mapsto  \omega^{ab}(Z)=dZ^M \omega^{ab}_M(Z)=E^A\omega^{ab}_A\; , \qquad E^A:= (E^a,  {\cal E} )=dZ^ME_M^A(Z) \, , \qquad \\  \nonumber \\
%  A(x)=e^aA_a \mapsto A(Z)=E^A A_A(Z)\; ,  \qquad  F_{ab}(x)\mapsto  F_{ab}(Z)\;  \qquad
%\nonumber \end{eqnarray}
%
\begin{equation}
  \label{e->E}
  \begin{array}{rcrcl}
    e^{a} &\longmapsto & E^{a} & =& dZ^{M}\ E_{M}{}^{a}(Z) \\
    \psi  &\longmapsto & \mathcal{E} & =& dZ^{M}\mathcal{E}_{M}(Z) = \sqrt{2}(E_{\alpha}^{ i},\bar{E}^{\dot{\alpha}}_{i})\\[.3cm]
    & \implies & E^{A} & :=& (E^a,  {\cal E} ) \ =\ dZ^{M} E_{M}{}^{A}(Z)\\[.3cm]
    \omega^{ab} &\longmapsto & \omega^{ab}(Z) & =& dZ^{M}\omega_{M}{}^{ab}(Z) \ =\ E^{C}\omega_{C}{}^{ab} \\
    A &\longmapsto & A(Z) & =& E^{C} A_{C}(Z) \\
    F_{ab}(x) &\longmapsto & F_{ab}(Z) & &
  \end{array}
\end{equation}
which implies that the Lorentz- and SU(2)-invariant bilinears are lifted as
\begin{eqnarray}
  \label{JmZ=}
 J_m\longmapsto  J_m(Z)
 & =&
 i\bar{{\cal E}}\wedge \tau^2 {\cal E}
 \; =\;
 2E^{\alpha i}\wedge E_{\alpha}^j \epsilon_{ij}- 2\bar{E}_{\dot{\alpha} i}\wedge \bar{E}{}_j^{\dot{\alpha}}\epsilon^{ij}
 \; , \\[.2cm]
 % \nonumber \\
 \tilde{J}_e  \; \longmapsto \;   \tilde{J}_e (Z)
 & =&
 -(\bar{{\cal E}}\wedge \gamma_5\tau^2\wedge {\cal E})
 \; =\; + 2i (E^{\alpha i}\wedge E_{\alpha}^j \epsilon_{ij}+\bar{E}_{\dot{\alpha} i}\wedge \bar{E}{}_j^{\dot{\alpha}}\epsilon^{ij})
 \; ,
\end{eqnarray}
where we added the expressions in terms of Weyl spinor fermionic vielbein
forms with explicit SU(2) indices for the reader's convenience.

Besides the above mentioned lifting of the differential forms to superspace,
passing to the generalized action implies replacing the integration over
spacetime $M_4$ with the integration over a surface ${\cal M}_4$ in
superspace, that is defined parametrically by fermionic coordinate functions
$\theta (x)$, i.e.~by $\theta =\theta (x)$.

If no independent equations result under the variation $\delta\theta (x)$ of
$\theta (x)$ (i.e.~from the variation of the surface ${\cal M}_4$), which is
the case for the above discussed generalized action of ${\cal N}=2$
supergravity (see \cite{Castellani:1991eu} for its equivalent form), one can
then lift the equations obtained from the generalized action to whole
superspace.\footnote{ The equations obtained from the generalized actions
  are functions on ${\cal M}_4$, i.e.~with $\theta$ substituted by
  $\theta (x)$ in the arguments of the superforms and the superfields.  } This
stage is the essence of the {\it rheonomic} or group manifold approach to
supergravity \cite{Neeman:1978njh,Castellani:1991eu},\footnote{ Two comments
  are in order: the first one is that the consistency of the lifted structure
  is not guaranteed and has to be checked.  The second comment is that our
  presentation of the rheonomic approach does not coincide literally with its
  original formulation \cite{Neeman:1978njh,Castellani:1991eu}: rather, we
  used \cite{Bandos:1995dw}'s reformulation, which generalizes the rheonomic
  approach to superstrings and supermembranes.  } and results in the lift of
all the differential-form equations of motion (\ref{Ta=N2x}--\ref{RSeq=N2M})
to superspace.  From the point of view of the superspace approach to
supergravity, (the formal algebraic solutions of) these lifted equations of
motion encode both the superspace constraints and their consequences,
including the superfield equations of motion of ${\cal N}=2$ supergravity.

%%%%%%%%%%%%%%%%%%%%%%%%%%%%%%%%%%%%%%%%%%%%%%%%%%%%%%%%%%%%%%%%%%%%%%
%%%%%%%%%%%%%%%%%%%%%%%%%%%%%%%%%%%%%%%%%%%%%%%%%%%%%%%%%%%%%%%%%%%%%%
%%%%%%%%%%%%%%%%%%%%%%%%%%%%%%%%%%%%%%%%%%%%%%%%%%%%%%%%%%%%%%%%%%%%%%
%%%%%%%%%%%%%%%%%%%%%%%%%%%%%%%%%%%%%%%%%%%%%%%%%%%%%%%%%%%%%%%%%%%%%%
\section{On-shell superspace of minimal ${\cal N}=2$ supergravity}
%%%%%%%%%%%%%%%%%%%%%%%%%%%%%%%%%%%%%%%%%%%%%%%%%%%%%%%%%%%%%%%%%%%%%%
%%%%%%%%%%%%%%%%%%%%%%%%%%%%%%%%%%%%%%%%%%%%%%%%%%%%%%%%%%%%%%%%%%%%%%
%%%%%%%%%%%%%%%%%%%%%%%%%%%%%%%%%%%%%%%%%%%%%%%%%%%%%%%%%%%%%%%%%%%%%%
%%%%%%%%%%%%%%%%%%%%%%%%%%%%%%%%%%%%%%%%%%%%%%%%%%%%%%%%%%%%%%%%%%%%%%

The superspace constraints for minimal ${\cal N}=2$ supergravity, which as
described above can be obtained from the generalized action, imply
\begin{eqnarray}\label{Ta=N2}
  T^a
  & =&
  DE^a
  \ =\ \frac i 2 \bar{{\cal E}}\wedge \gamma^a{\cal E}
  \; , \\
  % & & \nonumber  \\
  \label{Tf=N2M}
  {\cal T}
  & =&
  D{\cal E}
  \ =\
  \frac 1 8 E^a\wedge \slashed{F}\gamma_a\tau^2{\cal E}
  + \frac 1 2 E^b\wedge E^a\ {\cal T}_{ab}
  \; ,\qquad \slashed{F} \equiv F^{bc}\gamma_{bc}
  \; , \\
  %
  %& & \nonumber \\
  %
  \label{Rab=SSP}
  R^{ab}
  & :=&
  d\omega^{ab} -\omega^{ac}\wedge \omega_c{}^{b}
  \nonumber \\
  % &&\nonumber \\
  & =&
  -\frac i 4\bar{{\cal E}} (F^{ab}+\gamma_5*F^{ab})\wedge \tau^2  {\cal E} -i E^c \bar{{\cal T}}{}^{ab}\gamma_c\wedge {\cal E}
  +\frac 1 2 E^d\wedge E^c   R_{cd}{} ^{ab}
  \; , \\
  % && \nonumber \\
  \label{dA=Je+F2Z}
  dA
  & =&
  J_m(Z)+F_2(Z)
  \; ,
\end{eqnarray}
where
\begin{equation}
  F_2(Z)
  \ :=\ \frac 1 2 E^b\wedge E^a\ F_{ab}(Z)
  , \quad
  *F_2(Z)
  \ = \ \frac 14 E^b\wedge E^a\ \epsilon_{abcd} F^{cd}(Z)
  \; .
\end{equation}
%\begin{eqnarray} && F_2(Z):= \frac 1 2 E^b\wedge E^a F_{ab}(Z), \quad
%*F_2(Z)=\frac 14 E^b\wedge E^a \epsilon_{abcd} F^{cd}(Z)\; .  \qquad
%\end{eqnarray}

These are {\it on-shell constraints} because their consistency conditions
(i.e.~the superspace Bianchi identities) require the superfield
generalizations of the graviton, the gravitino and the gauge field strengths
to obey the superfield generalizations of the Einstein, Rarita-Schwinger and
Maxwell equations:
\begin{eqnarray}
  0 & =&
  \label{EiEq(Z)=}
  R_{ab}{}^{c  b}(Z) -\frac 1 2 \delta_{a}{}^{c} R_{de}{}^{de}(Z)
  - F_{ab}F^{bc} - \tfrac{1}{4} \delta_{a}{}^{c} F_{de}F^{de}(Z)
  \; ,\\
  \label{RS(Z)=N2}
  0 & =&
  {\cal T}_{[ab}\gamma_{c]}
  \;\; ,\hspace{4cm}
  {\cal T}_{ab} \; =\; \tfrac{i}{2}\ \epsilon_{abcd} \gamma_5 {\cal T}^{cd}
  \; , \\[.2cm]
  0
  & =&
  D^{b}F_{ab}(Z)
  \hspace{1.2cm} \Longleftrightarrow \hspace{1.2cm}
  d(*F_2(Z)-\tilde{J}_e(Z))=0
  \; .
\end{eqnarray}
%\begin{eqnarray}
%\label{EiEq(Z)=}
%R_{ab}{}^{c  b}(Z) -\frac 1 2 \delta_{a}{}^{c} R_{de}{}^{de}(Z)= F_{ab}F^{bc} +\frac 1 4 \delta_{a}{}^{c} F_{de}F^{de}(Z)
%\; , \qquad
%\\ \nonumber
%\\ \label{RS(Z)=N2}
%{\cal T}_{[ab}\gamma_{c]}=0 \; , \qquad{\cal T}_{ab} = \frac i 2 \epsilon_{abcd} \gamma_5 {\cal T}^{cd} \; , \qquad
%\\
%\nonumber
%\\
%  D^bF_{ab}(Z)=0\qquad \Leftrightarrow \qquad d(*F_2(Z)-\tilde{J}_e(Z))=0 \; . \qquad \\ \nonumber
%\end{eqnarray}

Let us repeat that all these equations can be obtained by lifting the
spacetime supergravity equations in their form given in equations
(\ref{Ta=N2x}--\ref{RSeq=N2M}) to superspace, which is tantamount to saying
that they are derived from the generalized action principle described in the
previous section.

Notice that the supersymmetry transformations (\ref{N2=susy},\ref{omN2=susy})
and \eqref{*F2=susy} can be obtained from the superspace constraints
(\ref{Ta=N2}--\ref{dA=Je+F2Z}) and/or from the superspace generalization of
the equations (\ref{Ta=N2x}--\ref{RSeq=N2M}) by the method based on the Lie
derivative formula which was described in the Appendix of
\cite{Bandos:2023zbs}.

%%%%%%%%%%%%%%%%%%%%%%%%%%%%%%%%%%%%%%%%%%%%%%%%%%%%%%%%%%%%%%%%%%%%%%
%%%%%%%%%%%%%%%%%%%%%%%%%%%%%%%%%%%%%%%%%%%%%%%%%%%%%%%%%%%%%%%%%%%%%%
%%%%%%%%%%%%%%%%%%%%%%%%%%%%%%%%%%%%%%%%%%%%%%%%%%%%%%%%%%%%%%%%%%%%%%
%%%%%%%%%%%%%%%%%%%%%%%%%%%%%%%%%%%%%%%%%%%%%%%%%%%%%%%%%%%%%%%%%%%%%%
\section{Killing supervector of ${\cal N}=2$ supergravity and the super-Komar form}
\label{superKomar}
%%%%%%%%%%%%%%%%%%%%%%%%%%%%%%%%%%%%%%%%%%%%%%%%%%%%%%%%%%%%%%%%%%%%%%
%%%%%%%%%%%%%%%%%%%%%%%%%%%%%%%%%%%%%%%%%%%%%%%%%%%%%%%%%%%%%%%%%%%%%%
%%%%%%%%%%%%%%%%%%%%%%%%%%%%%%%%%%%%%%%%%%%%%%%%%%%%%%%%%%%%%%%%%%%%%%
%%%%%%%%%%%%%%%%%%%%%%%%%%%%%%%%%%%%%%%%%%%%%%%%%%%%%%%%%%%%%%%%%%%%%%

A Killing supervector of ${\cal N}=2$ supergravity
\cite{Buchbinder:1995uq,Bandos:2023zbs}
%(i.e. of supergravity without gauging of $R$-symmetry group)
%
\begin{eqnarray}
  K^A
  \; =\;
  (K^a, {\cal K})
  \;\; ,\;\;
  {\cal K}
  \; =\;
  \sqrt{2}\left( \ K_\alpha^i \ ,\  K_i^{\dot{\alpha}} \right)
  \; ,
\end{eqnarray}

\noindent
is defined by the condition that the corresponding superspace
diffeomorphism (superdiffeomorphism) with parameters $K^A(Z)$ leaves all the
basic superfields and superforms inert {\it up to gauge transformations}, {\em
  i.e.\/} up to a $K$-dependent local Lorentz transformation with parameters
${L}_{K(Z)}^{ab}(Z)$ and up to a $K$-dependent $U(1)$ gauge symmetry
transformation with parameter $\chi_{K}(Z)$. This means that
\begin{eqnarray}
  \label{DKA=sKilling}
  -\delta_K E^A
  & =&
  DK^A + \imath_K T^A + E^B \imath_{K}{\omega}_B{}^A
  =
       E^B {L}_{K(Z)\; B}{}^A
       \nonumber \\
  & & \nonumber \\
  & =: &
  {\small
  \left(\begin{matrix}E^b{L}_{K(Z)\; b}{}^a & 0 \cr 0 & -\frac 1 4 {L}_{K(Z)}{}^{cd} \gamma_{cd}{\cal E}\end{matrix}\right)
  }
  \\
  %& & \nonumber \\
  \label{iKR=sKilling}
  -\delta_K \omega^{ab}
  & =&
  D\imath_{K}\omega^{ab} + \imath_{K} R^{ab}
  =
  D {L}_{K(Z)}^{ab}
  \; ,\\
  \label{iKdA=Killing}
  -\delta_K A
  & =&
  \imath_{K} (dA) + d\imath_{K}A
  =
  d\chi_{K}(Z)
  \; .
\end{eqnarray}
%\noindent
%{\it and} $U(1)$ gauge symmetry with $K$-dependent parameter $\chi_{_K}(Z)$,
%
%\bea\label{iKdA=Killing}
%-\delta_K A
%=
%\imath_{K} (dA) + d\imath_{K}A
%=
%d\chi_{K}(Z)
%\; .
%\eea

Using the constraints of ${\cal N}=2$ supergravity, we obtain the following
set of equations for the Killing supervector
\begin{eqnarray}
  \label{DKa=N2M}
  DK^a
  & =&
  -i \bar{\cal E}\gamma^a {\cal K}
  + E^b P_{(K)b}{}^a
  \, , \qquad
  \bar{{\cal E}}
  =
  -\sqrt{2}\left( \begin{matrix} E{}^{\alpha i}
  \, , \,
  \bar{E}{}_{\dot{\alpha}i} \end{matrix}\right)
  =
  dZ^M \bar{{\cal E}}_M(Z)
  \; ,\\
  %& & \nonumber \\
  \label{DKal=N2M}
  D {\cal K}
  & =&
  - \frac 1 8 E^a F\!\!\!/{}\, \gamma_a \tau^2 {\cal K}
  + \frac 1 8 K^a F\!\!\!\!/{}\, \gamma_a \tau^2 {\cal E}
  - E^c K^b {\cal T}_{bc}{}
  -  \frac 1 4  P_{(K)}^{ab}{\gamma}_{ab}\ {\cal E}
  \; ,
\end{eqnarray}
and the definition of the superfield Lorentz momentum map
\begin{equation}
  P_{(K)}{}^{ab}
  =
  -\imath_{K}{\omega}{}^{ab}+ {L}{}^{ab} (K)\,,
\end{equation}

\noindent
and the superfield 'electric' momentum map associated to the $U(1)$ gauge
symmetry

\begin{equation}
  P_{(K)}
  =
  -\imath_{K}A+ \chi_{(K)}\,.
  \, ,
\end{equation}

These momentum maps satisfy the equations

\begin{eqnarray}
\label{DPKab=N2M}
  DP_{(K)}^{ab}
  & =&
  \imath_{K}R^{ab}
  =
  -\tfrac{i}{2}\ \bar{{\cal E}} (F^{ab}+\gamma_5*F^{ab})\tau^2  {\cal K}
  -i E^c \bar{{\cal T}}{}^{ab}\gamma_c{\cal K}
  +  E^dK^c   R_{cd}{} ^{ab}
  \; , \\
  % &&  \nonumber\\
  \label{DPKel=N2M}
  dP_{(K)}
  & =&
  \imath_{K}dA
  \dot{=}
  \imath_{K}(F_2+J_m(Z))
  =
  2i\bar{{\cal E}}\tau^2 {\cal K}
  + E^bK^a   F_{ab}(Z)
  \; .
 \end{eqnarray}

 \noindent
 These last two equations can be also obtained from the consistency of the
 super-Killing equations \eqref{DKa=N2M} and \eqref{DKal=N2M}.

 However, this is not the end of story: somewhat surprisingly, we need to
 introduce also the 'magnetic' momentum map \cite{Ortin:2022uxa}, which has
 nothing to do with consistency of the above super-Killing equations.  This
 magnetic momentum map obeys
\begin{eqnarray}
  \label{dtPZ=M}
  d\tilde{P}_K(Z)
  & =&
  \imath_{K}\left(*F_2(Z)-\tilde{J}_e(Z)\right)
  =
  \tfrac{1}{2}\ K^a E^b\epsilon_{abcd}F^{cd} + 2\bar{{\cal E}} \gamma_5\tau^2{\cal K}
  \; .
\end{eqnarray}
The consistency condition of this equation, namely
$d\left(\imath_K(*F_2-\tilde{J}_e)\right)\doteq 0$, can be obtained from the
requirement that the tensor superfield $F_{ab}(Z)$ (and, consequently, the
superform $*F_2$) be invariant under the super-Killing transformation. In this
way of obtaining \eqref{dtPZ=M}, the superfield equation
$d(*F_2(Z)-\tilde{J}_e(Z))=0$ has to be used.

The magnetic momentum map $\tilde{P}_K(Z)$ is strictly necessary to construct
the
%superspace generalization of the Komar charge 2-form,
super-Komar 2-form,
i.e.~a 2-superform in superspace ${\mathbf{Q}}_2(\delta_{K(Z)})$, that is closed on-shell:
\begin{equation}
  \label{dQ2KZ=0}
  d{\mathbf{Q}_{2}}(\delta_{K(Z)})
  \; \doteq\;
  0
  \; .
\end{equation}

Following the construction of \cite{Bandos:2023zbs} and taking into account
the magnetic momentum map, one finds that the super-Komar form for minimal
$\mathcal{N}=2$, $d=4$ supergravity is given by
\begin{eqnarray}
  \label{Q2=KZ=Komar}
  {\mathbf{Q}}_2(\delta_{K(Z)})
  & =&
  2\ \bar{{\cal K}}\gamma \wedge {\cal E}
  +\tfrac{1}{2}\ E^a\wedge E^b\ \epsilon_{abcd} P^{cd}_{K}(Z)\
  \nonumber \\[.2cm]
  &&
  -\tfrac{1}{2}\ P_{K}(Z)(*F_2-\tilde{J}_e(Z))
  \ +\
  \tfrac{1}{2}\ \tilde{P}_{K}(Z)\, (F_2(Z)+J_m(Z))
  \; .
\end{eqnarray}

The closure of the super-Komar 2-form in on-shell superspace guarantees that
its leading component
\begin{eqnarray}
  \label{Q2kkap=}
  {\mathbf{Q}}_2(\delta_{(k,\kappa )})
  & \; :=\; &
              \left. {\mathbf{Q}} _2(\delta_{K(Z)})\right|_{\theta=0}
              \nonumber \\
  & & \nonumber \\
  & =&
  2\ \bar{\kappa}\gamma \wedge \psi
  -* {\cal P}_{2\, (k,\kappa) }
  -\tfrac{1}{2} {\cal P}_{(k,\kappa )}(*F_2-\tilde{J}_e)
  +\tfrac{1}{2} \tilde{{\cal P}}_{(k,\kappa )}dA
  %\nonumber\\
  %\nonumber  \\
  %& \doteq&
  %2\bar{\epsilon}\gamma \wedge \psi
  %-* {\cal P}_{2\, (k,\kappa)}
  %-\frac 1 2  {\cal P}_{k,\kappa}(*F_2-\tilde{J}_e)
  %+\frac 1 2  \tilde{{\cal P}}_{k,\kappa}(F_2+J_m)
  \; ,
\end{eqnarray}
is a closed spacetime 2-form
\begin{equation}
  d{\mathbf{Q}}_2(\delta_{(k,\kappa )})
  \; \doteq\;
  0
  \; ,
\end{equation}
and that it is, up to a total derivative, invariant under supersymmetry:
\begin{eqnarray}
  \label{susyQ2=}
  \delta_\epsilon {\mathbf{Q}}_2(\delta_{(k,\kappa)} )
  & =&
  d(\imath_\epsilon {\mathbf{Q}}_2(\delta_{(k,\kappa)}))
  \; , \\
  \nonumber \\
  \label{ieQ2=}
  \imath_\epsilon {\mathbf{Q}}_2(\delta_{(k,\kappa)})
  & =&
  2\bar{{\kappa}}\gamma {\epsilon}
  + i\bar{\psi}(\tilde{{\cal P}}_{(k,\kappa)}
  +i\gamma_5  {\cal P}_{(k,\kappa)})\tau^2 \ \epsilon
  \; .
\end{eqnarray}

This identifies the expression in \eqref{Q2kkap=} as the generalized Komar
2-form of minimal ${\cal N}=2$ supergravity with the desired
properties.\footnote{ Even though we have not dealt with the question of how
  chiral-duality transformations are implemented in the on-shell superspace, a
  comparison with the results in appendix \ref{sec:Appendix} shows that the
  generalized Komar 2-form (\ref{Q2kkap=}) is in fact invariant under
  chiral-duality transformations.  } In it
${\cal P}^{ab}_{(k,\kappa)}= {\cal P}^{ab}_{K(Z)}\vert_{\theta =0}$ and
${\cal P}_{(k,\kappa )}={\cal P}_{K(Z)}\vert_{\theta =0}$ are the Lorentz and
the 'electric' momentum maps defined as the leading components of
\eqref{DPKab=N2M} and \eqref{DPKel=N2M};
$\tilde{{\cal P}}_{(k,\kappa )}=\tilde{{\cal P}}_{K(Z)}\vert_{\theta =0}$ is
the 'magnetic' momentum map defined as the leading component of
\eqref{dtPZ=M}, i.e.~by the equation
\begin{eqnarray}
  \label{tcp:= v*F-J+}
  d\tilde{{\cal P}}_{(k,\kappa )}
  \ =\
  \imath_{k}(*F_2-\tilde{J}_e)
  -\imath_{\kappa-\imath_k\psi}\tilde{J}_e
  \ =\
  \imath_{k}*F_{2}\ -\ \imath_{\kappa}\tilde{J}_{e}
  \ =\
  \imath_{k}*F_{2}\ +\ \bar{\psi}\gamma_{5}\tau^{2}\kappa
  \; .
\end{eqnarray}

As a way of finishing this section, let us mention that Simon's generalized
Komar form \cite{Simon:1984qb} is, up to a total derivative term and specific
choices for the various momentum maps, equal to the one given in
\eqref{Q2kkap=}.

%%%%%%%%%%%%%%%%%%%%%%%%%%%%%%%%%%%%%%%%%%%%%%%%%%%%%%%%%%%%%%%%%%%%%%
%%%%%%%%%%%%%%%%%%%%%%%%%%%%%%%%%%%%%%%%%%%%%%%%%%%%%%%%%%%%%%%%%%%%%%
%%%%%%%%%%%%%%%%%%%%%%%%%%%%%%%%%%%%%%%%%%%%%%%%%%%%%%%%%%%%%%%%%%%%%%
%%%%%%%%%%%%%%%%%%%%%%%%%%%%%%%%%%%%%%%%%%%%%%%%%%%%%%%%%%%%%%%%%%%%%%
\section{On the spacetime component derivation of the generalized Komar 2-form
  and the advantage of the superspace approach}
%%%%%%%%%%%%%%%%%%%%%%%%%%%%%%%%%%%%%%%%%%%%%%%%%%%%%%%%%%%%%%%%%%%%%%
%%%%%%%%%%%%%%%%%%%%%%%%%%%%%%%%%%%%%%%%%%%%%%%%%%%%%%%%%%%%%%%%%%%%%%
%%%%%%%%%%%%%%%%%%%%%%%%%%%%%%%%%%%%%%%%%%%%%%%%%%%%%%%%%%%%%%%%%%%%%%
%%%%%%%%%%%%%%%%%%%%%%%%%%%%%%%%%%%%%%%%%%%%%%%%%%%%%%%%%%%%%%%%%%%%%%

Let us begin by observing that the generalized Komar 2-form \eqref{Q2kkap=} is
a linear combination of Noether charge 2-forms \cite{Wald:1993nt}
corresponding to the local symmetries of the supergravity action and its
equations of motion, namely
\begin{eqnarray}
  \label{Q2=Komar}
  {\mathbf{Q}} _2(\delta_{(k,\kappa )} )
  & =&
  {\mathbf{Q}} _2(\delta_{\epsilon= \kappa} )
  + {\mathbf{Q}}_2(\delta_{L^{ab}
  =
  - {\cal P}^{ab}_{k,\kappa}})
  + {\mathbf{Q}}_2(\delta_{\Lambda=-\frac 1 2 {\cal P}_{k,\kappa}} )
  + {\mathbf{Q}}_2(\delta_{\tilde{\Lambda }
  =
  \frac 1 2{\tilde{\cal P}}_{(k,\kappa )}} )
  \nonumber \\
  \nonumber \\
  & =&
  2\bar{\kappa}\gamma \wedge \psi
  + \frac 1 2 \epsilon_{abcd} {\cal P}^{ab}_{(k,\kappa )}\  e^c\wedge e^d
  -\frac 1 2  {\cal P}_{(k,\kappa )}(*F_2-\tilde{J}_e)
  +\frac 1 2  \tilde{{\cal P}}_{(k,\kappa )}dA
  \; .
\end{eqnarray}

Here the second term
${\mathbf{Q}}_2(\delta_{L^{ab}}) = -\frac 1 2 \epsilon_{abcd} L^{ab}\
e^c\wedge e^d $ is the Noether charge 2-form for local Lorentz symmetry of the
supergravity action

\begin{equation}
  \label{Lorentz=N2}
  \delta_Le^a=e^bL_b{}^a
  \; , \qquad
  \delta_L \psi
  =
  -\frac 1 4  L^{ab}\gamma_{ab}\psi
  \,  \, , \qquad
  \delta_L\omega^{ab}
  =
  DL^{ab}
  \; ;
\end{equation}
the third term,
${\mathbf{Q}}_2(\delta_{\Lambda} )= \Lambda \, (*F_2-\tilde{J}_e)$, is the
Noether charge 2-form for the $U(1)$ gauge symmetry, which only acts on the
vector field of the supergravity multiplet as $ \delta_{\Lambda}A=d\Lambda$.
We can call this 2-form 'electric' as the fourth term in \eqref{Q2=Komar}
reads ${\mathbf{Q}}_2(\delta_{\tilde{\Lambda }})=\tilde{\Lambda } dA$, which
is associated to the gauge transformations of the dual potential (which is, of
course, only defined on-shell); in this sense the charge
${\mathbf{Q}}_2(\delta_{\tilde{\Lambda }})$ is called 'magnetic'.

Finally, the first term
${\mathbf{Q}} _2(\delta_{\epsilon} )=2\bar{\epsilon}\gamma \wedge \psi $ is
the (2-form associated to the) supercharge, i.e.~the Noether charge 2-form for
local supersymmetry (\ref{N2=susy},\ref{omN2=susy},\ref{*F2=susy}).  The
exterior derivative of this form gives the 3-form that is dual to the Noether
current for local supersymmetry

\begin{equation}
  {\mathbf{J}}_3(\delta_\epsilon )
  \doteq
  d {\mathbf{Q}} _2(\delta_\epsilon )
  \; .
\end{equation}

As we will see in appendix \ref{sec:Appendix}, the standard method in the
component formalism is much more cumbersome, although we have used inputs of
it in the above procedure.  This suggests that the generalization of the
superspace method will also be more efficient in the construction of the
generalized Komar 2-forms for more complicated higher ${\cal N}\leq 8$ and
higher dimensional supergravity theories.

%%%%%%%%%%%%%%%%%%%%%%%%%%%%%%%%%%%%%%%%%%%%%%%%%%%%%%%%%%%%%%%%%%%%%%
%%%%%%%%%%%%%%%%%%%%%%%%%%%%%%%%%%%%%%%%%%%%%%%%%%%%%%%%%%%%%%%%%%%%%%
%%%%%%%%%%%%%%%%%%%%%%%%%%%%%%%%%%%%%%%%%%%%%%%%%%%%%%%%%%%%%%%%%%%%%%
%%%%%%%%%%%%%%%%%%%%%%%%%%%%%%%%%%%%%%%%%%%%%%%%%%%%%%%%%%%%%%%%%%%%%%
\section{Conclusion and outlook}
%%%%%%%%%%%%%%%%%%%%%%%%%%%%%%%%%%%%%%%%%%%%%%%%%%%%%%%%%%%%%%%%%%%%%%
%%%%%%%%%%%%%%%%%%%%%%%%%%%%%%%%%%%%%%%%%%%%%%%%%%%%%%%%%%%%%%%%%%%%%%
%%%%%%%%%%%%%%%%%%%%%%%%%%%%%%%%%%%%%%%%%%%%%%%%%%%%%%%%%%%%%%%%%%%%%%
%%%%%%%%%%%%%%%%%%%%%%%%%%%%%%%%%%%%%%%%%%%%%%%%%%%%%%%%%%%%%%%%%%%%%%

In this article we constructed an on-shell closed, supersymmetric invariant
2-superform in on-shell minimal $\mathcal{N}=2$, $d=4$ superspace, that we
referred to as the super-Komar form. The super-Komar form's lowest component
gives a gauge, chiral-duality and supersymmetry invariant generalized Komar
2-form, and as such generalizes the one found by Simon \cite{Simon:1984qb}.
\par
Seeing the generic ingredients that were used in the construction, we have
argued that the most convenient method to search for generalized Komar 2-forms
in more complex versions of supergravity would be
\begin{itemize}
\item To introduce the Killing supervector and momentum maps in on-shell
  superspace,
\item To search for a closed 2-form ($(D-2)$--form in case $D\not= 4$) in the
  on-shell superspace that is constructed out of the Killing supervector and
  the superfield momentum maps, taking as the starting point the superspace
  generalization of the Noether-Wald-type charge 2-forms associated to the
  manifest local superspace symmetries.
\item As our experience with minimal ${\cal N}=2$ supergravity has shown, this
  procedure will require the introduction of further momentum maps that
  correspond to non-manifest (gauge) symmetries of the (superspace) equations
  of motion.  In ${\cal N}=2$ supergravity case this was given by the
  'magnetic' momentum map corresponding to gauge symmetry of a dual vector
  potential.
  \par
  The form of the contribution of these further momentum maps to the candidate
  super-Komar form should, however, be fixed by the action of the non-manifest
  symmetries on the superfields and the known momentum maps. Also, should the
  generalized Komar 2-form for the purely bosonic sector of a supergravity be
  known, this will greatly help in finding the superfield form of these
  further momentum maps, as well as in fixing the coefficients in the
  candidate super-Komar form. Actually, this form is essentially known for all
  4-dimensional, ungauged supergravities
  \cite{Mitsios:2021zrn,Ballesteros:2023iqb}.
\end{itemize}

With respect to the association between magnetic momentum maps and
non-manifest gauge symmetries in the superspace description, we would like to
make the following comment: in \cite{Bandos:2024pns}, the authors had a
preliminary look at the action of Killing supervectors and the introduction of
superfield momentum maps in on-shell $d=11$ supergravity (M-sugra) superspace
along the lines presented in this article, with the aim of constructing the
corresponding super-Komar 9-form.  The need to introduce an electric
2-superform momentum map is manifest from the traditional on-shell
supergravity description \cite{Cremmer:1980ru,Brink:1980az} and a magnetic
5-superform momentum map is also needed, as follows from the bosonic sector
\cite{Ortin:2022uxa}.\footnote{ For prior work on M-theory thermodynamics, see
  e.g.~\cite{Townsend:2001rg,Haas:2014spa,Dias:2019wof}.  }

Even though the need for the magnetic superfield 5-form momentum map is
non-manifest in the traditional superspace approach to on-shell 11D
supergravity, it becomes natural in its duality-symmetric description
elaborated by Candiello and Lechner in \cite{Candiello:1993di}: they showed
that in 11D supergravity the generic set-up of curved supergeometry described
by the constraints on torsion and curvature allows for the existence of not
only a closed 4-superform $\mathcal{F}_{4}(Z)$, serving as a field strength
for a 3--superform gauge potential, but also of an {\em a priori} independent
7-superform $\mathcal{F}_{7}(Z)$ that satisfies the Bianchi identity
$d\mathcal{F}_{7}=\mathcal{F}_{4}\wedge\mathcal{F}_{4}$ which imply the
existence of a dual 6-form potential\footnote{The existence of such a dual
  6-form potential was actually noticed in \cite{DAuria:1982uck} devoted to
  group manifold approach and hidden gauge symmetries of 11D supergravity, see
  also \cite{Bandos:2004xw,Bandos:2004ym} for further studies of these.}. The
investigation of Bianchi identities (which simplifies essentially in such a
duality--symmetric formulation) indicates that the 7-th rank antisymmetric
tensor superfield enclosed in ${\mathcal F}_7$ is Hodge dual to the 4-th rank
antisymmetric tensor superfield in ${\mathcal F}_4$\footnote{this latter was
  introduced already in the original papers
  \cite{Cremmer:1980ru,Brink:1980az}}. Thus the duality is not imposed but
appeared as an inevitable result of the construction of superforms.

%%%%%%%%%%%%%%%%%%
\if{}

Even though the need for the magnetic superfield 5-form momentum map is
non-manifest in the traditional superspace approach to on-shell M-sugra, it
becomes natural in a different description, namely in the Free Differential
Algebra approach to M-sugra, see e.g.~\cite{Castellani:1991eu}: in the
traditional approach, it was already known that the solutions to the
supertorsion constraints allow for the construction of a closed 4-superform
field strength $\mathcal{F}_{4}(Z)$ \cite{Cremmer:1980ru,Brink:1980az} and also
a 7-superform field strength $\mathcal{F}_{7}(Z)$, that satisfies the Bianchi
identity $d\mathcal{F}_{7}=\mathcal{F}_{4}\wedge\mathcal{F}_{4}$
\cite{Candiello:1993di}; by construction, the lowest component of the
7-superform is related to the Hodge dual of the lowest component of the
4-superform, i.e.~$\mathcal{F}_{7}(x)=\star\mathcal{F}_{4}(x)$. In
this way, the introduction superfield momentum maps associated to the action
of the Killing supervector is a convenient way of formulating the non-manifest
information contained in the super-Killing equations and the torsion
constraints. In the FDA approach to M-sugra, however, it was shown that a
direct rheonomic solution of the Bianchi identities governing the various
curvatures, that include independent 4- and 7-form superfield strengths, leads
to on-shell M-sugra \cite{Castellani:1991eu}. Reformulated in terms of
superspace, see e.g.~\cite{Bandos:2004ym}, we have that on-shell
M-sugra is equivalent to imposing the conventional constraints and the
existence of independent superforms $\mathcal{F}_{4}(Z)$ and
$\mathcal{F}_{7}(Z)$, that satisfy the aforementioned Bianchi identities; the
fact that $\mathcal{F}_{7}(x)=\star\mathcal{F}_{4}(x)$ is a consequence of the
construction, and as such the FDA description gives a democratic formulation
\cite{Cremmer:1998px} of on-shell M-sugra. In this democratic description, the
need to introduce a magnetic superspace momentum map associated to
$\mathcal{F}_{7}(Z)$, is as natural as the introduction of an electric
momentum map associated to $\mathcal{F}_{4}(Z)$.

\fi %%%%%%%%%%%

It stands to reason that a similar reformulation exists for on-shell minimal
$\mathcal{N}=2$, $d=4$ superspace and in such a democratic description, the
the magnetic momentum maps would arise most naturally.

% In the appendix \ref{appsec:FirstLaw}, we gave detailed the deduction of the
% 2-form that gives to the first law It is clearly a nice idea to find a
% superspace implementation of how to deduce the first-law form, and the
% details are currently being worked out.

In order to really show that the construction method found in this work is
really the most efficient one to find generalized Komar forms in supergravity,
more work needs to be done, as well as the implications of having an on-shell
superform instead of an on-shell form needs to be clarified. In the
introduction, we mentioned the possibility of doing supersymmetric
thermodynamics for the known black hole solutions in minimal $\mathcal{N}=2$,
$d=4$ supergravity with fermionic hair
\cite{Aichelburg:1983ux,Aichelburg:1984ef}.  Prior investigations into this
topic, see
e.g.~\cite{Simon:1984qb,Prabhu:2015vua,Aneesh:2020fcr,Aichelburg:1983ux},
would indicate that fermionic hair does not play any relevant role. This prior
work ignores, however, the compensating supersymmetry transformations, and we
are currently revisiting the topic.

%%%%%%%%%%%%%%%%%%%%%%%%%%%%%%%%%%%%%%%%%%%%%%%%%%%%%%%%%%%%%%%%%%%%%%
%%%%%%%%%%%%%%%%%%%%%%%%%%%%%%%%%%%%%%%%%%%%%%%%%%%%%%%%%%%%%%%%%%%%%%
%%%%%%%%%%%%%%%%%%%%%%%%%%%%%%%%%%%%%%%%%%%%%%%%%%%%%%%%%%%%%%%%%%%%%%
%%%%%%%%%%%%%%%%%%%%%%%%%%%%%%%%%%%%%%%%%%%%%%%%%%%%%%%%%%%%%%%%%%%%%%
\acknowledgments
%%%%%%%%%%%%%%%%%%%%%%%%%%%%%%%%%%%%%%%%%%%%%%%%%%%%%%%%%%%%%%%%%%%%%%
%%%%%%%%%%%%%%%%%%%%%%%%%%%%%%%%%%%%%%%%%%%%%%%%%%%%%%%%%%%%%%%%%%%%%%
%%%%%%%%%%%%%%%%%%%%%%%%%%%%%%%%%%%%%%%%%%%%%%%%%%%%%%%%%%%%%%%%%%%%%%
%%%%%%%%%%%%%%%%%%%%%%%%%%%%%%%%%%%%%%%%%%%%%%%%%%%%%%%%%%%%%%%%%%%%%%

This work has been supported in part by

\begin{enumerate}
\item The MCI, AEI, FEDER (UE) grant PID2021-125700NB-C21 (“Gravity,
  Supergravity and Superstrings” (GRASS)) (IB and TO).
\item The Basque Government Grant IT1628-22 (IB).
\item The MCI, AEI, FEDER (UE) grant PID2021-123021NB-I00 (PM).
\item FICYT through the Asturian grant SV-PA-21-AYUD/2021/52177 (PM).
\item The MCI, AEI, FEDER (UE) grant IFT Centro de Excelencia Severo Ochoa
  CEX2020-001007-S (TO).

\end{enumerate}

TO wishes to thank M.M.~Fern\'andez for her permanent support.

%%%%%%%%%%%%%%%%%%%%%%%%%%%%%%%%%%%%%%%%%%%%%%%%%%%%%%%%%%%%%%%%%%%%%%
%%%%%%%%%%%%%%%%%%%%%%%%%%%%%%%%%%%%%%%%%%%%%%%%%%%%%%%%%%%%%%%%%%%%%%
%%%%%%%%%%%%%%%%%%%%%%%%%%%%%%%%%%%%%%%%%%%%%%%%%%%%%%%%%%%%%%%%%%%%%%
%%%%%%%%%%%%%%%%%%%%%%%%%%%%%%%%%%%%%%%%%%%%%%%%%%%%%%%%%%%%%%%%%%%%%%
\appendix
%%%%%%%%%%%%%%%%%%%%%%%%%%%%%%%%%%%%%%%%%%%%%%%%%%%%%%%%%%%%%%%%%%%%%%
%%%%%%%%%%%%%%%%%%%%%%%%%%%%%%%%%%%%%%%%%%%%%%%%%%%%%%%%%%%%%%%%%%%%%%
%%%%%%%%%%%%%%%%%%%%%%%%%%%%%%%%%%%%%%%%%%%%%%%%%%%%%%%%%%%%%%%%%%%%%%
%%%%%%%%%%%%%%%%%%%%%%%%%%%%%%%%%%%%%%%%%%%%%%%%%%%%%%%%%%%%%%%%%%%%%%
  \section{The spacetime component approach}
  \label{sec:Appendix}
%%%%%%%%%%%%%%%%%%%%%%%%%%%%%%%%%%%%%%%%%%%%%%%%%%%%%%%%%%%%%%%%%%%%%%
%%%%%%%%%%%%%%%%%%%%%%%%%%%%%%%%%%%%%%%%%%%%%%%%%%%%%%%%%%%%%%%%%%%%%%
%%%%%%%%%%%%%%%%%%%%%%%%%%%%%%%%%%%%%%%%%%%%%%%%%%%%%%%%%%%%%%%%%%%%%%
%%%%%%%%%%%%%%%%%%%%%%%%%%%%%%%%%%%%%%%%%%%%%%%%%%%%%%%%%%%%%%%%%%%%%%
%%%%%%%%%
%
%%%% Change of notation   :: Done step-by-step as to avoid problems
%%
%%%%%%  Normalisation of slash :: DONE
%%%%%%  \Psi -> \psi  :: DONE
%%%%%%  \rho -> 2\Psi :: DONE
%%%%%%  \mathbb{R}_{ab} -> R_{ab}  :: DONE
%%%%%%  \aleph -> 1 :: DONE
%%%%%%  \mathbb{} -> \mathbf{}   :: DONE
%%%%%%  \Gamma -> \gamma_{(1)} :: DONE
%%%%%%  \mathcal{P}_{(0)}[k] -> \tilde{\mathcal{K}}_{k}  :: DONE
%%%%%%  \tilde{P}_{(0)}[k] -> \mathcal{P}_{k}   :: DONE
%%%%%%  Get rid of the labels indicating the degree of the forms
%
%%%%%%%%%%%%%%%%%%%%%%%%%%%%%%%%%%%%%%%%%%%%%%%%%%%%%%%%%%%%%%%%%%%%%%%%%%%%%%%%%%%%%%%%

  In this appendix we will derive the generalized Komar 2-form using the
  component formulation. This might be useful as a reference for people not
  accustomed to superspace techniques and it will also highlight the
  effectiveness of the superspace approach, as well as an independent
  confirmation of our main results.
\par
Contrary to the main part of the text, and more conventionally, in this
appendix we will have the (covariant) exterior derivative and the interior
product act from the left. Furthermore, we will use abbreviation such as
$e^{ab}=e^{a}\wedge e^{b}$ and will define the spin connection and Lorentz
covariant derivative by \cite{Bandos:2023zbs,Ortin:2015hya}
\begin{eqnarray}
  \mathcal{D}e^{a}
  & =&
  de^{a}\ -\ \omega^{a}{}_{b}\wedge e^{b}
  \; =\; -T^{a} \; , \\
  R_{a}{}^{b}
  & =&
  d\omega_{a}{}^{b}\ -\ \omega_{a}{}^{c}\wedge\omega_{c}{}^{b} \; ,\\
  \mathcal{D}\psi
  & =&
  d\psi \ -\ \tfrac{1}{4}\slashed{\omega}\wedge\psi
  \; =\; 2\ \Psi
  \; .
\end{eqnarray}
We define $A$'s field strength by $F=dA$, the fermionic bilinear 2-forms as
\begin{equation}
  \label{eq:63}
  \left.
  \begin{array}{lcl}
    S & =& i\ \overline{\psi}\wedge\sigma_{2}\psi \\[.1cm]
    % &  & \\
    C & =& \overline{\psi}\wedge \gamma_{5}\sigma_{2}\psi
  \end{array}
  \right\}
  \;\xrightarrow{\mbox{and the modified field strengths}}\;
  \left\{
    \begin{array}{ccc}
      \tilde{F} & =& F\ +\ S \\[.1cm]
      % & & \\
      \star\mathcal{F} & =& \star\tilde{F}\ +\ C
    \end{array}
  \right.
\end{equation}
where $\tilde{F}$ is the supercovariant field strength; in the main part of
this article the supercovariant fieldstrenth is denoted by $F_{2}$, and taking
into account the difference of conventions, we have $\tilde{F}=-F_{2}$.
%\begin{align}
%  \label{eq:63}
%  \mathcal{J}_{e}
%  & =
%    i\ \overline{\psi}\wedge \sigma_{2}\psi
%    \;\;\;\xrightarrow{\;\mbox{redefine}\;}\;\;\;
%    S_{(2)}
%    \; =\;
%    \mathcal{J}_{e}
%    \; =\;
%    i\ \overline{\psi}\wedge \sigma_{2}\psi
%    \; , \\
%  \mathcal{J}_{m}
%  & =
%    -\star\left( \overline{\psi}\wedge\gamma_{5}\sigma_{2}\psi\right)
%    \;\;\;\xrightarrow{\;\mbox{redefine}\;}\;\;\;
%    C_{(2)}
%    \; =\;
%    \star\mathcal{J}_{m}
%    \; =\;
%    \overline{\psi}\wedge \gamma_{5}\sigma_{2}\psi\; ,\\
%  \widetilde{F}
%  & =
%    F\ +\ \mathcal{J}_{e}
%    \; =\;
%    F\ +\ S_{(2)}
%    \; , \\
%  \mathcal{F}
%  & =
%    \widetilde{F}
%    \ +\
%    \mathcal{J}_{m}
%    \;\;\;\xrightarrow{\;\mbox{redefine}\;}\;\;\;
%    \star\mathcal{F}
%    \; =\;
%    \star\tilde{F}
%    \ +\
%    C_{(2)}
%    \; .
%\end{align}
\par
The Lagrangian top form is then written in the first-order formalism
as\footnote{ This action is first order in the sense that the spin-connection
  $\omega_{ab}$ is an independent field.  }
\begin{equation}
  \label{eq:108}
  \mathbf{L}
  \; = \;
  -\star e^{ab}\wedge {R}_{ab}
  \ +\
  2\overline{\psi}\wedge\gamma_{5}\gamma\wedge\mathcal{D}\psi
  \ +\
  \tfrac{1}{2} \widetilde{F}\wedge \star\widetilde{F}
  \ +\ \widetilde{F}\wedge C
  \ -\
  \tfrac{1}{2}\ S\wedge C
  \; .
\end{equation}
The equation of motion of the spin-connection $\omega_{ab}$ implies that the
torsion is determined as
\begin{equation}
  \label{eq:24}
  T^{a}
  \; =\;
  \tfrac{i}{2}\
  \overline{\psi}\wedge \gamma^{c}\psi
  %\;\;\;\;\;\mbox{so that}\;\;\;\;\;
  %T_{ab}{}^{c}
  %\; =\;
  %i\ \overline{\psi}_{a}\gamma^{c}\psi_{b}
  \; ,
\end{equation}
and in the spirit of the 1.5 order formalism, we'll consider this equation
solved when convenient.
\par
Under a generic variation of the three independent fields, we have
\begin{equation}
  \label{eq:21}
  \delta\mathbf{L}
  \; =\;
  \mathbf{E}_{a}\wedge \delta e^{a}
  %\ +\
  %\tfrac{1}{2}\ \mathbf{E}_{ab}\wedge\delta\omega^{ab}
  \ +\
  \mathbf{E}_{A}\wedge \delta A
  \ +\
  \overline{\delta\psi}\wedge\mathbf{E}
  \ +\
  d\mathbf{\Theta} (\phi ,\delta\phi )
  \; ,
\end{equation}
we find the surface contribution
\begin{align}
  \label{eq:23}
  \mathbf{\Theta} (\phi ,\delta\phi )
  & =\;
  -\star e^{ab}\wedge\delta\omega_{ab}
  \ +\
  \star\mathcal{F}\wedge\delta A
  \ -\
  2\ \overline{\delta\psi}\wedge\gamma_{5}\gamma\wedge\psi
  \; ,
\end{align}
and the Equation of Motion-forms
\begin{eqnarray}
  \label{eq:22}
  \mathbf{E}_{a}
  & =&
    {R}(\omega )_{bc}\wedge \imath_{a}\star e^{bc}
    \ +\
    2\ \overline{\psi}\wedge \gamma_{5}\gamma_{a}\mathcal{D}\psi
    \ +\
    \tfrac{1}{2}\left[
       \imath_{a}\tilde{F}\wedge \star\tilde{F}
       -
       \tilde{F}\wedge\imath_{a}\star\tilde{F}
    \right]
    \; , \\
  \mathbf{E}_{A}
  & =&
    -d\star\mathcal{F}
    \; , \\
  \label{eq:57}
  \mathbf{E}
  & =&
    4\gamma_{5}\gamma\wedge\hat{\mathcal{D}}\psi
    \; ,
\end{eqnarray}
where we defined the generalized covariant derivative
\begin{equation}
  \label{eq:71}
  \hat{\mathcal{D}}\psi
  \; \equiv\;
  \mathcal{D}\psi
  \ +\
  \tfrac{1}{8}\ \slashed{\widetilde{F}}\ \gamma\wedge\sigma_{2}\psi
  %\; =\;
  %2\ \hat{\Psi}
  \; .
\end{equation}
The above Lagrangian top form is strictly invariant under spacetime
diffeomorphisms, U(1) gauge transformation and local Lorentz transformations
as well as global SU(2) R-symmetry transformations; under local supersymmetry
transformations, however, it is invariant up to a total derivative.

%%%%%%%%%%%%%%%%%%%%%%%%%%%%%%%%%%%%%%%%%%%%%%%%%%%%%%%%%%%%%%%%%%%%%%
%%%%%%%%%%%%%%%%%%%%%%%%%%%%%%%%%%%%%%%%%%%%%%%%%%%%%%%%%%%%%%%%%%%%%%
%%%%%%%%%%%%%%%%%%%%%%%%%%%%%%%%%%%%%%%%%%%%%%%%%%%%%%%%%%%%%%%%%%%%%%
%%%%%%%%%%%%%%%%%%%%%%%%%%%%%%%%%%%%%%%%%%%%%%%%%%%%%%%%%%%%%%%%%%%%%%
\subsection*{The on-shell chiral-duality transformations}
%%%%%%%%%%%%%%%%%%%%%%%%%%%%%%%%%%%%%%%%%%%%%%%%%%%%%%%%%%%%%%%%%%%%%%
%%%%%%%%%%%%%%%%%%%%%%%%%%%%%%%%%%%%%%%%%%%%%%%%%%%%%%%%%%%%%%%%%%%%%%
%%%%%%%%%%%%%%%%%%%%%%%%%%%%%%%%%%%%%%%%%%%%%%%%%%%%%%%%%%%%%%%%%%%%%%
%%%%%%%%%%%%%%%%%%%%%%%%%%%%%%%%%%%%%%%%%%%%%%%%%%%%%%%%%%%%%%%%%%%%%%

As is well-known, chiral-duality in an on-shell symmetry of $\mathcal{N}=2$
$d=4$ supergravity and in our conventions it corresponds to the
simultaneous transformations
\begin{equation}
  \label{eq:156}
  \psi^{\prime}
  \; =\;
  e^{\frac{i}{2}\theta \gamma_{5}}\ \psi
  \;\;\; ,\;\;\;
  \tilde{F}^{\prime}
  \; =\;
  \cos (\theta )\ \tilde{F}
  \ +\
  \sin (\theta )\ \star\tilde{F}
  \; .
  \; ,
\end{equation}
so that $(\tilde{F},\star\tilde{F})$ transform as a doublet.  The chirality
transformation on the gravitino field, then implies that $(-S,C)$ transforms
as a doublet, which, together with the definitions in eq.~(\ref{eq:63}), implies
that $(F,\star\mathcal{F})$ also transforms as a doublet.
\par
Observe that $d\star\mathcal{F}\doteq 0$, so that we can
introduce a 1-form field $B$ such that locally $\star\mathcal{F}\doteq dB$:
the pair $(A,B)$ transforms as a doublet under chiral-duality transformations.
Furthemore, notice that $B$ depends on the fermions.

%%%%%%%%%%%%%%%%%%%%%%%%%%%%%%%%%%%%%%%%%%%%%%%%%%%%%%%%%%%%%%%%%%%%%%
%%%%%%%%%%%%%%%%%%%%%%%%%%%%%%%%%%%%%%%%%%%%%%%%%%%%%%%%%%%%%%%%%%%%%%
%%%%%%%%%%%%%%%%%%%%%%%%%%%%%%%%%%%%%%%%%%%%%%%%%%%%%%%%%%%%%%%%%%%%%%
%%%%%%%%%%%%%%%%%%%%%%%%%%%%%%%%%%%%%%%%%%%%%%%%%%%%%%%%%%%%%%%%%%%%%%
\subsection*{The local Lorentz Noether charge 2-form}
%%%%%%%%%%%%%%%%%%%%%%%%%%%%%%%%%%%%%%%%%%%%%%%%%%%%%%%%%%%%%%%%%%%%%%
%%%%%%%%%%%%%%%%%%%%%%%%%%%%%%%%%%%%%%%%%%%%%%%%%%%%%%%%%%%%%%%%%%%%%%
%%%%%%%%%%%%%%%%%%%%%%%%%%%%%%%%%%%%%%%%%%%%%%%%%%%%%%%%%%%%%%%%%%%%%%
%%%%%%%%%%%%%%%%%%%%%%%%%%%%%%%%%%%%%%%%%%%%%%%%%%%%%%%%%%%%%%%%%%%%%%

Under a local Lorentz transformation with position dependent parameters
$\sigma_{ab}=-\sigma_{ba}$, the fields transform as
\begin{equation}
  \label{eq:26}
  \delta_{\sigma}e^{a}
  \; =\;
  \sigma^{a}{}_{b}\ e^{b}
  \;\;\; ,\;\;\;
  \delta_{\sigma}\psi
  \; =\;
  \tfrac{1}{4}\
  \slashed{\sigma}\ \psi
  \;\;\;\mbox{and}\;\;\;
  \delta_{\sigma}\omega^{ab}
  \; =\;
  \mathcal{D}\sigma^{ab}
  \; .
\end{equation}
The Noether identity for local Lorentz transformations states that
\begin{equation}
  \label{eq:28}
  0
  \; =\;
  \mathbf{E}_{a}\wedge \delta_{\sigma}e^{a}
  \ +\
  \overline{\delta_{\sigma}\psi}\wedge\mathbf{E}
  \; ,
\end{equation}
which implies that the Noether 3-form $\mathbf{J}[\sigma]$ is off-shell
closed, whence locally exact, and is given by
\begin{equation}
  \label{eq:37}
  \mathbf{J}[\sigma]
  \; =\;
  \mathbf{\Theta} (\phi ,\delta_{\sigma}\phi)
  \; =\;
  d\mathbf{Q}[\sigma]
  \;\;\;\;\mbox{with}\;\;\;\;
  \mathbf{Q}[\sigma ]
  \; =\;
  -\star e^{ab} \wedge \sigma_{ab}
  \; .
\end{equation}
$\mathbf{Q}[\sigma ]$ is the Noether charge 2-form associated to the local
Lorentz transformation.

%%%%%%%%%%%%%%%%%%%%%%%%%%%%%%%%%%%%%%%%%%%%%%%%%%%%%%%%%%%%%%%%%%%%%%
%%%%%%%%%%%%%%%%%%%%%%%%%%%%%%%%%%%%%%%%%%%%%%%%%%%%%%%%%%%%%%%%%%%%%%
%%%%%%%%%%%%%%%%%%%%%%%%%%%%%%%%%%%%%%%%%%%%%%%%%%%%%%%%%%%%%%%%%%%%%%
%%%%%%%%%%%%%%%%%%%%%%%%%%%%%%%%%%%%%%%%%%%%%%%%%%%%%%%%%%%%%%%%%%%%%%
\subsection*{The U(1) Noether charge 2-form}
%%%%%%%%%%%%%%%%%%%%%%%%%%%%%%%%%%%%%%%%%%%%%%%%%%%%%%%%%%%%%%%%%%%%%%
%%%%%%%%%%%%%%%%%%%%%%%%%%%%%%%%%%%%%%%%%%%%%%%%%%%%%%%%%%%%%%%%%%%%%%
%%%%%%%%%%%%%%%%%%%%%%%%%%%%%%%%%%%%%%%%%%%%%%%%%%%%%%%%%%%%%%%%%%%%%%
%%%%%%%%%%%%%%%%%%%%%%%%%%%%%%%%%%%%%%%%%%%%%%%%%%%%%%%%%%%%%%%%%%%%%%

The action is strictly invariant under the local $U(1)$ transformation
%\begin{equation}
%  \label{eq:93}
%  \delta_{\chi}A
%  \; =\;
%  d\chi
%  \; ,
%\end{equation}
$\delta_{\chi}A=d\chi$, and using $d\mathbf{E}_{A}=0$, we immediately find the
electric Noether charge 2-form $\mathbf{Q}[\chi ]$ to be
\begin{equation}
  \label{eq:94}
  \mathbf{J}[\chi ]
  \; =\;
  \mathbf{\Theta} (\phi ,\delta_{\chi}\phi )
  \ -\
  \chi\ \mathbf{E}_{A}
  \; =\;
  d\mathbf{Q}[\chi ]
  \;\;\;\;\mbox{with}\;\;\;\;
  \mathbf{Q}[\chi ]
  \; =\;
  \chi\ \star\mathcal{F}
  \; .
\end{equation}
%being the electric Noether charge-form.

%%%%%%%%%%%%%%%%%%%%%%%%%%%%%%%%%%%%%%%%%%%%%%%%%%%%%%%%%%%%%%%%%%%%%%
%%%%%%%%%%%%%%%%%%%%%%%%%%%%%%%%%%%%%%%%%%%%%%%%%%%%%%%%%%%%%%%%%%%%%%
%%%%%%%%%%%%%%%%%%%%%%%%%%%%%%%%%%%%%%%%%%%%%%%%%%%%%%%%%%%%%%%%%%%%%%
%%%%%%%%%%%%%%%%%%%%%%%%%%%%%%%%%%%%%%%%%%%%%%%%%%%%%%%%%%%%%%%%%%%%%%
\subsection*{The supersymmetry Noether charge 2-form}
%%%%%%%%%%%%%%%%%%%%%%%%%%%%%%%%%%%%%%%%%%%%%%%%%%%%%%%%%%%%%%%%%%%%%%
%%%%%%%%%%%%%%%%%%%%%%%%%%%%%%%%%%%%%%%%%%%%%%%%%%%%%%%%%%%%%%%%%%%%%%
%%%%%%%%%%%%%%%%%%%%%%%%%%%%%%%%%%%%%%%%%%%%%%%%%%%%%%%%%%%%%%%%%%%%%%
%%%%%%%%%%%%%%%%%%%%%%%%%%%%%%%%%%%%%%%%%%%%%%%%%%%%%%%%%%%%%%%%%%%%%%

Even though the supersymmetry transformations are well known, in the end we
will be interested in on-shell supersymmetry including the on-shell 1-form
field $B$: its supersymmetry transformation can be deduced by demanding
closure of the supersymmetry algebra and compatibility with the chiral-duality
transformations \cite{Huebscher:2010ib}.  The resulting supersymmetry
transformations are
\begin{align}
  \label{eq:46}
  \delta_{\epsilon}e^{a}
  & =
    -i\ \bar{\epsilon}\gamma^{a}\psi
    \; , \\
  \label{eq:46b}
  \delta_{\epsilon}A
  & =
    -2i\ \bar{\epsilon}\sigma_{2}\psi
    \; , \\
  \label{eq:46d}
  \delta_{\epsilon}B
  & =
  2\ \bar{\epsilon}\gamma_{5}\sigma_{2}\psi
  \; ,\\
  \label{eq:46c}
  \delta_{\epsilon}\psi
  & =
    \hat{\mathcal{D}}\epsilon
    \; =\;
    \mathcal{D}\epsilon
    \ +\
    \tfrac{1}{8} \slashed{\widetilde{F}}\ \gamma\
    \sigma_{2}\epsilon
    \; .
\end{align}
Under an explicit supersymmetry transformation, the Lagrangian top form
transforms as $\delta_{\epsilon}\mathbf{L}=d\mathbf{F}_{\epsilon}$, with
\begin{equation}
  \label{eq:55}
  \mathbf{F}_{\epsilon}
  \;  =\;
  -\star e^{ab}\wedge\delta_{\epsilon}\omega_{ab}
  \; +2\ \bar{\epsilon}\gamma_{5}\gamma\wedge \mathcal{D}\psi
  \; +\
  F\wedge \bar{\epsilon}\gamma_{5}\sigma_{2}\psi
  \; -\
  i\ \star\mathcal{F}\wedge \bar{\epsilon}\sigma_{2}\psi
  \; .
\end{equation}
The corresponding Noether 3-form is then derived by using the susy-Noether
identity
\begin{equation}
  \label{eq:SusyNoether}
  \mathcal{D}\mathbf{E}
  +
  \tfrac{1}{8}\gamma\wedge \slashed{\tilde{F}}\sigma_{2}\mathbf{E}
  \; =\;
  -i\mathbf{E}_{a}\wedge \gamma^{a}\psi
  \ -\
  2i
  \mathbf{E}_{A}\wedge \sigma_{2}\psi
  \; .
\end{equation}
The Noether 3-form associated to supersymmetry is then
\begin{equation}
  \label{eq:56}
  \mathbf{J}[\epsilon ]
  \; =\;
  \mathbf{\Theta} (\delta_{\epsilon}\phi )
  \ -\
  \mathbf{F}_{\epsilon}
  \ +\
  \bar{\epsilon}\ \mathbf{E}
  \; =\;
  d\mathbf{Q}[\epsilon ]
  \;\;\;\;\; \mbox{with}\;\;\;
  \mathbf{Q}[\epsilon ]
  \; =\;
  -2\ \bar{\epsilon}\gamma_{5}\gamma\wedge\psi
  \; ,
\end{equation}
where $\mathbf{Q}[\epsilon ]$ is the supersymmetry Noether charge 2-form.
%Even though this term is not the prettiest one, it is near trivial to
%see that it leads to
%\begin{equation}
%  \label{eq:121}
%  \mathbf{J}_{(3)}[\epsilon ]
%  \; =\; d\mathbf{Q}_{(2)}[\epsilon ]
%  \;\;\; \mbox{with}\;\;\;
%  \mathbf{Q}_{(2)}[\epsilon ]
%  \; =\;
%  -2\ \bar{\epsilon}\gamma_{5}\gamma\wedge\psi
%  \; .
%\end{equation}
%Observe that this is formally the same as the one in $N=1$ $d=4$ \cite{Bandos:2023zbs}.
\par
Following \cite{Barnich:2001jy,Barnich:2003xg}, we would like to find the
conditions under which the supersymmetry parameter $\epsilon$ is a
reducibility parameter, i.e.~the restrictions on $\epsilon$ such that
on-shell we have $d\mathbf{Q}[\epsilon ]\doteq 0$.  A straightforward
calculation using a Fierz identity to eliminate a contribution from the
torsion, gives:
% This was already checked
\begin{equation}
  \label{eq:RedCondition}
  0
  \ \doteq\
  d\mathbf{Q}[\epsilon ]
  \ \doteq\
  2\overline{\psi}\gamma_{5}\left(
    \gamma\wedge\mathcal{D}\epsilon
    \ +\
    \tfrac{1}{2}\ F\sigma_{2}\epsilon
    \ +\
    \tfrac{i}{2}\ \star\mathcal{F}\gamma_{5}\sigma_{2}\epsilon
    \; ,
  \right)
  \; .
\end{equation}
Equating the terms between the parenthesis to zero, we find that it is
equivalent to the following generalized Killing spinor equation
\begin{equation}
  \label{eq:SusyBarnich}
  0
  \; =\;
  \mathcal{D}\epsilon
  +\tfrac{1}{16}\left(\slashed{\mathcal{F}}+\slashed{F}\right)\gamma\sigma_{2}\epsilon
  +\tfrac{1}{48}\gamma \left(\slashed{\mathcal{F}}-\slashed{F}\right)\sigma_{2}\epsilon
  \; .
\end{equation}
This equation is chiral-duality covariant and reduces to the ordinary Killing
spinor equation $\hat{\mathcal{D}}\epsilon =0$ when $\psi=0$.

%%%%%%%%%%%%%%%%%%%%%%%%%%%%%%%%%%%%%%%%%%%%%%%%%%%%%%%%%%%%%%%%%%%%%%
%%%%%%%%%%%%%%%%%%%%%%%%%%%%%%%%%%%%%%%%%%%%%%%%%%%%%%%%%%%%%%%%%%%%%%
%%%%%%%%%%%%%%%%%%%%%%%%%%%%%%%%%%%%%%%%%%%%%%%%%%%%%%%%%%%%%%%%%%%%%%
%%%%%%%%%%%%%%%%%%%%%%%%%%%%%%%%%%%%%%%%%%%%%%%%%%%%%%%%%%%%%%%%%%%%%%
\subsection*{The Noether-Wald 2-form, definitions of the momentum maps}
%%%%%%%%%%%%%%%%%%%%%%%%%%%%%%%%%%%%%%%%%%%%%%%%%%%%%%%%%%%%%%%%%%%%%%
%%%%%%%%%%%%%%%%%%%%%%%%%%%%%%%%%%%%%%%%%%%%%%%%%%%%%%%%%%%%%%%%%%%%%%
%%%%%%%%%%%%%%%%%%%%%%%%%%%%%%%%%%%%%%%%%%%%%%%%%%%%%%%%%%%%%%%%%%%%%%
%%%%%%%%%%%%%%%%%%%%%%%%%%%%%%%%%%%%%%%%%%%%%%%%%%%%%%%%%%%%%%%%%%%%%%

Following the general strategy outlined in \cite{Elgood:2020svt}, we will
write the variation of a given field $\phi$ under an infinitesimal general
coordinate transformation (GCT) generated by the vector $\xi$, using the
covariant Lie derivative, that is to say: in terms of the field strengths and
momentum maps. These covariant Lie derivatives guarantee that if we choose the
GCT to be generated by a vector $k$ such that $\delta_{k}\phi =0$ for all the
fields, these conditions are gauge-covariant. We will call these vector fields
Killing vectors since, in particular, the GCTs they generate must leave
invariant the metric. We will denote them generically by $k$.
\par
Consider for example the 1-form field $A$: as the field transforms under its
gauge symmetry
%\eqref{eq:93}
$\delta_{\chi}A=d\chi$ and supersymmetry \eqref{eq:46b}, we can write the most
general variation as
\begin{equation}
  \label{eq:90}
  -\delta_{\xi}A
  =
  \pounds_{\xi}A
  - d\chi_{\xi}
  - \delta_{\epsilon_{\xi}}A
  =
  \imath_{\xi}\tilde{F}
  +
  d\mathcal{P}_{\xi}
  -2i\overline{\lambda}_{\xi}\sigma_{2}\psi
  \;\;\;\mbox{with}\;\;
  \left\{
    \begin{array}{lcl}
      \mathcal{P}_{\xi}
      & =&
           \imath_{\xi}A -\chi_{\xi}
      \\
      & & \\
      \lambda_{\xi}
      & =&
           \imath_{\xi}\psi
           -
           \epsilon_{\xi}
    \end{array}
    \right.
\end{equation}
where we defined the so-called electric momentum map $\mathcal{P}_{\xi}$. In
the same sense $\lambda_{\xi}$ could be called the fermionic momentum map.
\par
When $\xi=k$ (a Killig vector in the generalized sense that we have defined),
the condition of invariance $\delta_{k}A=0$ implies that
\begin{equation}
  \delta_{k}A \ =\ 0
  \;\;\;\implies\;\;\;
  \imath_{k}\tilde{F}
  \; =\;
  -d\mathcal{P}_{k}
  +
  2i\ \overline{\kappa}\sigma_{2}\psi
  \; ,
\end{equation}
where we renamed the fermionic momentum map as $\kappa \equiv \lambda_{k}$,
and see that this corresponds to the generalized Killing spinor.\footnote{ In the main part
  of the text the variation of the fields w.r.t.~the Killing vector $k$ is
  denoted by $\delta_{( k,\kappa ) }$ as the pair $(k ,\kappa)$ give the
  lowest order contribution of the Killing supervector $K^{A}$. As we will see
  in eqs.~(\ref{eq:600}--\ref{eq:600d}), $(k ,\kappa)$, $\mathcal{P}_{k}$
  and the magentic momentum map $\tilde{\mathcal{P}}_{k}$ that will be
  introduced next, form a vector supermultiplet. It would therefore be natural
  to label the variation of the fields using a reference to this
  supermultiplet, but in order to keep the notation as clean as possible, we
  will just write $\delta_{k}\phi$.  }
\par
For the on-shell $B$ field, taking into account its supersymmetry
transformation \eqref{eq:46d} and its gauge transformation
$\delta_{\tilde{\chi}} B=d\tilde{\chi}$, we find that
\begin{equation}
  \label{eq:90b}
  -\delta_{\xi}B
  \ =\
  \imath_{\xi}\star\mathcal{F}
  +
  d\tilde{\mathcal{P}}_{\xi}
  +
  2\ \overline{\lambda_{\xi}}\gamma_{5}\sigma_{2}\psi
  \;\;\; \mbox{where}\;\;\;
  \tilde{\mathcal{P}}_{\xi}
  =
  \imath_{\xi}B-\tilde{\chi}_{\xi}
  \; ,
\end{equation}
is given the name of magnetic momentum map \cite{Ortin:2022uxa}; clearly,
$(\mathcal{P}_{\xi},\tilde{\mathcal{P}}_{\xi})$ is a doublet under
chiral-duality transformations.
\par
There is one missing momentum map: it is found by observing that the vierbein
transforms under supersymmetry and under local Lorentz transformations. This
leads to
\begin{equation}
  \label{eq:91}
  -\delta_{\xi}e^{a}
  \ =\
  \mathcal{D}\xi^{a}
    +
    P_{\xi}^{a}{}_{b}e^{b}
    -
    i\overline{\lambda_{\xi}}\gamma^{a}\psi
  \;\;\;\mbox{where}\;\;
  P_{\xi}^{ab}
  \ =\
  \imath_{\xi}\omega^{ab}
  -
  \sigma_{\xi}^{ab}
  \; ,
\end{equation}
is called the Lorentz momentum map.  There are no further momentum maps that
can arise and we have
\begin{eqnarray}
  \label{eq:126}
  -\delta_{\xi}\omega_{ab}
  & =&
  \imath_{\xi} {R}_{ab}
  \; + \;
  \mathcal{D}P_{\xi ab}
  \ -\
  \delta_{\epsilon_{\xi}}\omega_{ab}
  \; ,\\
  \label{eq:92}
  -\delta_{\xi}\psi
  & =&
    \imath_{\xi}\hat{\mathcal{D}}\psi
    +
    \hat{\mathcal{D}}\lambda_{\xi}
    +
    \tfrac{1}{4}\ P_{\xi}^{ab}\gamma_{ab}\psi
    -
    \tfrac{1}{8}\ \slashed{\tilde{F}}\ \slashed{\xi}\ \sigma_{2}\psi
  \; .
\end{eqnarray}
A lengthy calculation using the GCT Noether identity
\begin{equation}
  0
  \; =\;
  \xi^{a}\mathcal{D}\mathbf{E}_{a}
  \ +\
  \mathbf{E}_{A}\wedge\imath_{\xi}\tilde{F}
  \ +\
  \tfrac{1}{8}\ \overline{\psi}\wedge \slashed{\xi}\ \slashed{\tilde{F}}\sigma_{2}\mathbf{E}
  \; ,
\end{equation}
leads to the mid-point result
\begin{equation}
  \label{eq:125}
  \mathbf{J}[\xi ]
  \ =\
  \mathbf{\Theta} (\delta_{\xi}\phi )
  +
  \xi^{a}\mathbf{E}_{a}
  +
  \imath_{\xi}\mathbf{L}
  +
  \mathcal{P}_{\xi}\mathbf{E}_{A}
  -
  \overline{\lambda_{\xi}}\mathbf{E}
  -
  \mathbf{F}_{\epsilon_{\xi}}
  \; ,
\end{equation}
where $\mathbf{F}_{\epsilon_{\xi}}$ is given in eq.~(\ref{eq:55}).
\par
Finally, after the mid-point result one rapidly finds the off-shell
Noether-Wald 2-form $\mathbf{Q}[\xi ]$ by
\begin{equation}
  \label{eq:127}
  \mathbf{J}[\xi ]
  \; =\;
  d\mathbf{Q}[\xi ]
  \;\;\;\mbox{with}\;\;\;
  \mathbf{Q}[\xi ]
  \; =\;
  2\ \star P_{2}[\xi ]
  \ -\
  \mathcal{P}_{\xi}\ \star\mathcal{F}
  \ +\
  2\ \overline{\lambda_{\xi}}\gamma_{5}\gamma\wedge \psi
  \; ,
\end{equation}
where we have defined the 2-form $P_{2}[\xi ] = \tfrac{1}{2}\ P_{\xi ab}\
e^{ab}$.

%Observe that the bosonic part coincides with the expression in ref.~\cite{Elgood:2020svt}
%and with the $N=1$ result in \cite{Bandos:2023zbs}.

%%%%%%%%%%%%%%%%%%%%%%%%%%%%%%%%%%%%%%%%%%%%%%%%%%%%%%%%%%%%%%%%%%%%%%
%%%%%%%%%%%%%%%%%%%%%%%%%%%%%%%%%%%%%%%%%%%%%%%%%%%%%%%%%%%%%%%%%%%%%%
%%%%%%%%%%%%%%%%%%%%%%%%%%%%%%%%%%%%%%%%%%%%%%%%%%%%%%%%%%%%%%%%%%%%%%
%%%%%%%%%%%%%%%%%%%%%%%%%%%%%%%%%%%%%%%%%%%%%%%%%%%%%%%%%%%%%%%%%%%%%%
\subsection*{The generalized Komar 2-form from the spacetime component approach}
%%%%%%%%%%%%%%%%%%%%%%%%%%%%%%%%%%%%%%%%%%%%%%%%%%%%%%%%%%%%%%%%%%%%%%
%%%%%%%%%%%%%%%%%%%%%%%%%%%%%%%%%%%%%%%%%%%%%%%%%%%%%%%%%%%%%%%%%%%%%%
%%%%%%%%%%%%%%%%%%%%%%%%%%%%%%%%%%%%%%%%%%%%%%%%%%%%%%%%%%%%%%%%%%%%%%
%%%%%%%%%%%%%%%%%%%%%%%%%%%%%%%%%%%%%%%%%%%%%%%%%%%%%%%%%%%%%%%%%%%%%%

If we consider the GCT to be generated by a Killing vector $k$ that leaves a
chosen solution invariant, whence $\delta_{k}\phi=0$, the 3-form in
eq.~(\ref{eq:125}) is such that \cite{Mitsios:2021zrn}
\begin{equation}
  \label{eq:122}
  \mathbf{J}[k]
  \; \doteq\;
  \imath_{k}\mathbf{L}
  \ -\
  \mathbf{F}_{\epsilon_{k}}
  \; =\;
  d\varpi_{k}
  \; ,
\end{equation}
where the last identification follows from the fact that $\mathbf{J}[k]$ must
be locally exact.
\par
Using Fierz identities and the equations of motion, we have
\begin{align}
 0 & =\;
 \overline{\psi}\wedge \gamma_{5}\mathcal{D}\gamma\wedge\psi
 \; ,\\
 \label{eq:135}
 \mathbf{L}
 & \doteq\;
 \tfrac{1}{2}\ F\wedge \star\mathcal{F}
 \; , \\
 \label{eq:501}
 \delta_{\epsilon}\omega_{ab}
 & =\;
 -2i\bar{\epsilon}\gamma\ \hat{\Psi}_{ab}
 \ -\
 \tfrac{i}{2}\ \tilde{F}_{ab}\ \bar{\epsilon}\sigma_{2}\psi
 \ +\
 \tfrac{1}{2}\
 \left(\star\tilde{F}\right)_{ab}\ \bar{\epsilon}\gamma_{5}\sigma_{2}\psi
 \; ,\\
 \label{eq:146}
 -\mathbf{F}_{\epsilon_{k}}
 & \doteq\;
 -F\wedge \bar{\epsilon}_{k}\gamma_{5}\sigma_{2}\psi
 \ +\
 i\star\mathcal{F}\wedge \bar{\epsilon}_{k}\sigma_{2}\psi
 \; .
\end{align}
Putting it together, we see that
\begin{align}
  d\varpi_{k}
  & \doteq\;
  \tfrac{1}{2}\left(
    \imath_{k}F
    +
    2i\bar{\epsilon}_{k}\sigma_{2}\psi
  \right)
  \wedge \star\mathcal{F}
  \ +\
  \tfrac{1}{2} F\wedge\left(
    \imath_{k}\star\mathcal{F}
    -
    2\bar{\epsilon}_{k}\gamma_{5}\sigma_{2}\psi
  \right)
  \nonumber \\
  & =
  d\left(
    -\tfrac{1}{2}\ \mathcal{P}_{k}\ \star\mathcal{F}
    \ -\
    \tfrac{1}{2}\ \tilde{\mathcal{P}}_{k}\ F
  \right)
  \; ,
\end{align}
where we used the definitions of the electric and magnetic momentum maps in
(\ref{eq:90}) and (\ref{eq:90b}).
\par
Defining the generalized Komar 2-form
$\mathbf{K}[k]=\varpi_{k}-\mathbf{Q}[k]$, where $\mathbf{Q}[k]$ is the GCT
Noether-Wald in (\ref{eq:127}) specified to the case $\xi=k$ and
$\lambda_{k}=\kappa$, we have
\begin{equation}
  \label{eq:162}
  \mathbf{K}[k]
  \; =\;
  -2\ \star P_{2}[k ]
  \ -\
  2\ \bar{\kappa}\gamma_{5}\gamma\wedge \psi
  \ +\
  \tfrac{1}{2}\ \mathcal{P}_{k}\ \star\mathcal{F}
  \ -\
  \tfrac{1}{2}\ \tilde{\mathcal{P}}_{k}\ F
  \; ,
\end{equation}
which by construction is on-shell closed, i.e.~$d\mathbf{K}[k]\doteq 0$, and is
furthermore invariant under chiral-duality transformations; it coincides, up
to conventions, with the generalized Komar form given in \eqref{Q2kkap=}.
\par
An $N=2$ $d=4$ vector multiplet consists of a vector, 2 Majorana spinors and 1
complex scalar, which is enough to accommodate the Killing vector $k^{a}$, the
generalized Killing spinors $\kappa$ and the electric and magnetic momentum
maps $(\mathcal{P}_{k},\tilde{\mathcal{P}}_{k})$. Taking clues from
\cite{Bandos:2023zbs}, the general superspace construction and chiral-duality
covariance, one finds that the supersymmetry variations
\begin{align}
  \label{eq:600}
  \delta_{\epsilon}k^{a}
  & =
  -i\ \bar{\epsilon}\gamma^{a}\kappa
  \; , \\
  \label{eq:600b}
  \delta_{\epsilon}\kappa
  & =
  -\tfrac{1}{4}\ \slashed{P}_{2}\epsilon
  \ +\
  \tfrac{1}{8}\ \slashed{\tilde{F}}\ \slashed{k}\ \sigma_{2}\epsilon
  \; ,\\
  \label{eq:600c}
  \delta_{\epsilon}\mathcal{P}_{k}
  & =
  -2i\ \bar{\epsilon}\sigma_{2}\kappa
  \; ,\\
  \label{eq:600d}
  \delta_{\epsilon}\tilde{\mathcal{P}}_{k}
  & =
  2\ \bar{\epsilon}\gamma_{5}\sigma_{2}\kappa
  \; .
\end{align}
lead to the standard closure of the $N=2$ superalgebra, when we take into
account that the $0$-form momentum maps $\mathcal{P}_{k}$ and
$\tilde{\mathcal{P}}_{k}$ are $A$- and $B$-gauge invariant. Hence, these
objects do fill a vector supermultiplet of $\mathcal{N}=2$, $d=4$
supergravity.
\par
Using the supersymmetry rules, the equations of motion and some Fierz
identities, one finds that
\begin{equation}
  \label{eq:Komar-SUSYvar}
  \delta_{\epsilon}\mathbf{K}
  \; \doteqdot \;
  \mathrm{d}\left[
    2\ \bar{\epsilon}\gamma_{5}\gamma\kappa
    +
    \mathcal{P}_{k}\ \bar{\epsilon}\gamma_{5}\sigma_{2}\psi
    +
    i\ \tilde{\mathcal{P}}_{k}\ \bar{\epsilon}\sigma_{2}\psi
  \right]
  \; ,
\end{equation}
so that the generalized Komar 2-form is, up to a total derivative, invariant
under supersymmetry.

%%%%%%%%%%%%%%%%%%%%%%%%%%%%%%%%%%%%%%%%%%%%%%%%%%%%%%%%%%%%%%%%%%%%%%
%%%%%%%%%%%%%%%%%%%%%%%%%%%%%%%%%%%%%%%%%%%%%%%%%%%%%%%%%%%%%%%%%%%%%%
%%%%%%%%%%%%%%%%%%%%%%%%%%%%%%%%%%%%%%%%%%%%%%%%%%%%%%%%%%%%%%%%%%%%%%
%%%%%%%%%%%%%%%%%%%%%%%%%%%%%%%%%%%%%%%%%%%%%%%%%%%%%%%%%%%%%%%%%%%%%%
\subsection{First law 2-form from the component approach}
\label{appsec:FirstLaw}
%%%%%%%%%%%%%%%%%%%%%%%%%%%%%%%%%%%%%%%%%%%%%%%%%%%%%%%%%%%%%%%%%%%%%%
%%%%%%%%%%%%%%%%%%%%%%%%%%%%%%%%%%%%%%%%%%%%%%%%%%%%%%%%%%%%%%%%%%%%%%
%%%%%%%%%%%%%%%%%%%%%%%%%%%%%%%%%%%%%%%%%%%%%%%%%%%%%%%%%%%%%%%%%%%%%%
%%%%%%%%%%%%%%%%%%%%%%%%%%%%%%%%%%%%%%%%%%%%%%%%%%%%%%%%%%%%%%%%%%%%%%

In this small subsection we would like to outline the construction of the
first law 2-form, along the lines of
\cite{Wald:1993nt,Prabhu:2015vua,Jacobson:2015uqa,Aneesh:2020fcr,Elgood:2020svt,Ortin:2022uxa,Bandos:2023zbs}:
starting out from the general definition of the presymplectic 3-form
\begin{equation}
  \label{eq:163}
  \Omega (\delta\phi ,\delta_{\xi}\phi )
  \; \doteq\;
  \delta\mathbf{\Theta} (\delta\phi ,\delta_{\xi}\phi )
  \ -\
  \delta_{\xi}\mathbf{\Theta} (\phi ,\delta\phi )
  \; ,
\end{equation}
where the variation $\delta\phi$ is between two solutions to the equations of
motion, and using standard manipulations using the various off-shell relations
\cite{Ortin:2022uxa}, we find that
\begin{equation}
  \label{eq:130}
  \Omega (\delta\phi ,\delta_{\xi}\phi )
  \; \doteq\;
  d\left[
    \delta\mathbf{Q}[\xi ]
    +
    \imath_{\xi}\mathbf{\Theta} (\delta\phi )
  \right]
  \ +\
  \delta\mathbf{F}_{\epsilon_{\xi}}
  \ -\
  \delta_{g(\xi )}\mathbf{\Theta} (\delta\phi )
  \; ,
\end{equation}
where $\delta_{g(\xi )}$ is the variation w.r.t.~the compensating gauge-,
local Lorentz- and supersymmetry transformations associated to the GCT
generated by $\xi$.
\par
For a GCT generated by the Killing vector $k$ that leaves all field invariant,
$\delta_{k}\phi=0$, the presymplectic form vanishes, so that locally we have
\begin{equation}
  0
  \; \doteq\;
  d\left[
    \delta\mathbf{Q}[k ]
    +
    \imath_{k}\mathbf{\Theta} (\delta\phi )
  \right]
  \ +\
  \delta\mathbf{F}_{\epsilon_{k}}
  \ -\
  \delta_{g(k )}\mathbf{\Theta} (\delta\phi )
  \; =\;
  d\mathbf{W}[k]
  \; ,
\end{equation}
where we'll refer to the 2-form $\mathbf{W}[k]$ as the first law 2-form.
\par
As the contributions of the compensating U(1) gauge- and local Lorentz
transformations to the first law 2-form are rather standard (see
e.g.~\cite{Ortin:2022uxa}), we will limit the discussion to the compensating
supersymmetry transformation with parameter $\epsilon_{k}$.
\par
Taking into account that the variation $\delta\phi$ is between solutions, one
has
\begin{align}
  \delta_{\epsilon_{k}}\delta A
    & \doteq\;
    -2i\ \overline{\epsilon_{k}}\sigma_{2}\delta\psi
    \ -\
    2i\ \overline{\delta\epsilon_{k}}\ \sigma_{2}\psi
    \; , \\
    \delta_{\epsilon_{k}}\delta\psi
    & \doteq\;
    \hat{\mathcal{D}}\delta\epsilon_{k}
    -\tfrac{1}{4}\delta\omega_{ab}\gamma^{ab}\epsilon_{k}
    +\tfrac{1}{8}\delta\tilde{F}_{ab} \gamma^{ab}\gamma\sigma_{2}\epsilon_{k}
    +\tfrac{1}{8}\slashed{\tilde{F}}\ \delta\gamma\ \sigma_{2}\epsilon_{k}
    \; ,\\
    \star e^{ab}\wedge \delta_{\epsilon_{k}}\delta\omega_{ab}
      & \doteq\;
      2\ \overline{\epsilon_{k}}\gamma_{5}\gamma\wedge \delta\left(\hat{\mathcal{D}}\psi \right)
      -\delta\left[
        \tilde{F}\wedge \overline{\epsilon_{k}}\gamma_{5}\sigma_{2}\psi
        + i\
        \star\tilde{F}\wedge \overline{\epsilon_{k}}\sigma_{2}\psi
      \right]
      \nonumber \\
      & \;\;\;\;
      +\delta e^{a}\wedge\left(
        \imath_{a}\tilde{F}\wedge\overline{\epsilon_{k}}\gamma_{5}\sigma_{2}\psi
        \ +\
        i\ \imath_{a}\star\tilde{F}\wedge \overline{\epsilon_{k}}\sigma_{2}\psi
      \right)
      \; ,
\end{align}
which means that as far as the supersymmetry contribution is concerned, we
have
\begin{equation}
  \label{eq:503}
    \delta\mathbf{F}_{\epsilon_{k}}-\delta_{\epsilon_{k}}\mathbf{\Theta} (\delta\phi)
    \; \doteq\;
    d\left[
      2\ \overline{\delta\epsilon_{k}}\gamma_{5}\gamma\wedge\psi
      -
      2\ \overline{\epsilon_{k}}\gamma_{5}\gamma\wedge\delta\psi
      -
      2\ \overline{\epsilon_{k}}\gamma_{5}\sigma_{2}\psi\wedge\delta A
    \right]
    \; .
\end{equation}
The end result then is that the first law 2-form is given by
\begin{eqnarray}
  \label{eq:505}
  \mathbf{W}[k]
  & = &
  P_{k ab}\ \delta\star e^{ab}
  - \mathcal{P}_{k}\ \delta\star\mathcal{F}
  + \tilde{\mathcal{P}}_{k}\ \delta F
  +\ \bar{\kappa}\ \delta\left( 2\gamma_{5}\gamma\wedge \psi\right)
  \nonumber \\
  & &
  \ +\
  2\bar{\kappa}\gamma_{5}\gamma\wedge\delta\psi
  -\imath_{k}\star e^{ab}\wedge \delta\omega_{ab}
  \ -\
  2\overline{\psi}\wedge\gamma_{5}\slashed{k}\ \delta\psi
  \; .
\end{eqnarray}
In the case of vanishing fermions the obtained first law 2-form corresponds to
the one for ordinary 4-dimensional Einstein-Maxwell gravity (see
e.g.~\cite{Ortin:2022uxa}). Even though the first law 2-form for minimal
$\mathcal{N}=1$, $d=4$ supergravity with active fermions was not calculated in
\cite{Bandos:2023zbs}, the authors calculated it before deriving the one
displayed in \eqref{eq:505}: the $\mathcal{N}=1$ result can be obtained
eliminating the gauge field contributions and truncating the spinors.

The physical interpretation of all the terms which are obtained after  integrating $\mathbf{W}[k]$ of \eqref{eq:505} over spheres at infinity and at the horizon is yet to be investigated.

\bibliographystyle{JHEP} %unsrt, iopart-num
\bibliography{ENTROSUGRAN2-V2.bib}

\providecommand{\href}[2]{#2}\begingroup\raggedright\begin{thebibliography}{10}

\bibitem{Komar:1958wp}
A.~Komar, \emph{{Covariant conservation laws in general relativity}},
  \href{https://doi.org/10.1103/PhysRev.113.934}{\emph{Phys. Rev.} {\bfseries
  113} (1959) 934}.

\bibitem{Arnowitt:1962hi}
R.L.~Arnowitt, S.~Deser and C.W.~Misner, \emph{{The Dynamics of general
  relativity}}, \href{https://doi.org/10.1007/s10714-008-0661-1}{\emph{Gen.
  Rel. Grav.} {\bfseries 40} (2008) 1997}
  [\href{https://arxiv.org/abs/gr-qc/0405109}{{\ttfamily gr-qc/0405109}}].

\bibitem{Abbott:1981ff}
L.F.~Abbott and S.~Deser, \emph{{Stability of Gravity with a Cosmological
  Constant}}, \href{https://doi.org/10.1016/0550-3213(82)90049-9}{\emph{Nucl.
  Phys. B} {\bfseries 195} (1982) 76}.

\bibitem{Ballesteros:2024prz}
R.~Ballesteros and T.~Ortin, \emph{{Generalized Komar charges and Smarr
  formulas for black holes and boson stars}},
  \href{https://arxiv.org/abs/2409.08268}{{\ttfamily 2409.08268}}.

\bibitem{Smarr:1972kt}
L.~Smarr, \emph{{Mass formula for Kerr black holes}},
  \href{https://doi.org/10.1103/PhysRevLett.30.71}{\emph{Phys. Rev. Lett.}
  {\bfseries 30} (1973) 71}.

\bibitem{Bardeen:1973gs}
J.M.~Bardeen, B.~Carter and S.W.~Hawking, \emph{{The Four laws of black hole
  mechanics}}, \href{https://doi.org/10.1007/BF01645742}{\emph{Commun. Math.
  Phys.} {\bfseries 31} (1973) 161}.

\bibitem{Carter:1973rla}
B.~Carter, \emph{{Black holes equilibrium states}},  in \emph{{Les Houches
  Summer School of Theoretical Physics}: {Black Holes}}, {DeWitt, C. and
  DeWitt, B.}, ed., pp.~57--214, 1973.

\bibitem{Simon:1984qb}
W.~Simon, \emph{{Gravitational field strengths and generalized Komar
  integral}}, \href{https://doi.org/10.1007/BF00761903}{\emph{Gen. Rel. Grav.}
  {\bfseries 17} (1985) 439}.

\bibitem{Kosmann:1971}
Y.~Kosmann, \emph{D\'eriv\'ees de lie des spineurs},
  \href{https://doi.org/10.1007/BF02428822}{\emph{Annali di Matematica}
  {\bfseries 91} (1971) 317}.

\bibitem{Vandyck:1988ei}
M.A.J.~Vandyck, \emph{{On the problem of space-time symmetries in the theory of
  supergravity}}, \href{https://doi.org/10.1007/BF00759185}{\emph{Gen. Rel.
  Grav.} {\bfseries 20} (1988) 261}.

\bibitem{Ortin:2002qb}
T.~Ort\'{\i}n, \emph{{A Note on Lie-Lorentz derivatives}},
  \href{https://doi.org/10.1088/0264-9381/19/15/101}{\emph{Class. Quant. Grav.}
  {\bfseries 19} (2002) L143}
  [\href{https://arxiv.org/abs/hep-th/0206159}{{\ttfamily hep-th/0206159}}].

\bibitem{Fatibene2011}
L.~Fatibene and M.~Francaviglia, \emph{General theory of lie derivatives for
  lorentz tensors}, {\emph{Communications in Mathematics} {\bfseries 19} (2011)
  11}.

\bibitem{Prabhu:2015vua}
K.~Prabhu, \emph{{The First Law of Black Hole Mechanics for Fields with
  Internal Gauge Freedom}},
  \href{https://doi.org/10.1088/1361-6382/aa536b}{\emph{Class. Quant. Grav.}
  {\bfseries 34} (2017) 035011}
  [\href{https://arxiv.org/abs/1511.00388}{{\ttfamily 1511.00388}}].

\bibitem{Elgood:2020svt}
Z.~Elgood, P.~Meessen and T.~Ort\'{\i}n, \emph{{The first law of black hole
  mechanics in the Einstein-Maxwell theory revisited}},
  \href{https://doi.org/10.1007/JHEP09(2020)026}{\emph{JHEP} {\bfseries 09}
  (2020) 026} [\href{https://arxiv.org/abs/2006.02792}{{\ttfamily
  2006.02792}}].

\bibitem{Vandyck:1988gc}
M.A.~Vandyck, \emph{{On the problem of space-time symmetries in the theory of
  supergravity. II: N=2 supergravity and spinorial Lie derivatives}},
  \href{https://doi.org/10.1007/BF00760090}{\emph{Gen. Rel. Grav.} {\bfseries
  20} (1988) 905}.

\bibitem{Bandos:2023zbs}
I.~Bandos and T.~Ort\'{\i}n, \emph{{Noether-Wald charge in supergravity: the
  fermionic contribution}},
  \href{https://doi.org/10.1007/JHEP12(2023)095}{\emph{JHEP} {\bfseries 12}
  (2023) 095} [\href{https://arxiv.org/abs/2305.10617}{{\ttfamily
  2305.10617}}].

\bibitem{Ortin:2022uxa}
T.~Ort\'{\i}n and D.~Pere\~n\'{i}guez, \emph{{Magnetic charges and Wald
  entropy}}, \href{https://doi.org/10.1007/JHEP11(2022)081}{\emph{JHEP}
  {\bfseries 11} (2022) 081}
  [\href{https://arxiv.org/abs/2207.12008}{{\ttfamily 2207.12008}}].

\bibitem{Ortin:2024mmg}
T.~Ort\'{\i}n and M.~Zatti, \emph{{A note on the Noether-Wald and generalized
  Komar charges}},  \href{https://arxiv.org/abs/2411.10420}{{\ttfamily
  2411.10420}}.

\bibitem{Buchbinder:1995uq}
I.L.~Buchbinder and S.M.~Kuzenko, \emph{{Ideas and methods of supersymmetry and
  supergravity: a walk through superspace}}, CRC Press (1998, revised edition).

\bibitem{Vandyck:1989ai}
M.A.~Vandyck, \emph{{On the problem of space-time symmetries in the theory of
  supergravity. III: superspace formalism}},
  \href{https://doi.org/10.1007/BF00756186}{\emph{Gen. Rel. Grav.} {\bfseries
  21} (1989) 79}.

\bibitem{Kuzenko:2012vd}
S.M.~Kuzenko, \emph{{Symmetries of curved superspace}},
  \href{https://doi.org/10.1007/JHEP03(2013)024}{\emph{JHEP} {\bfseries 03}
  (2013) 024} [\href{https://arxiv.org/abs/1212.6179}{{\ttfamily 1212.6179}}].

\bibitem{Kuzenko:2015lca}
S.M.~Kuzenko, \emph{{Supersymmetric spacetimes from curved superspace}},
  \href{https://doi.org/10.22323/1.231.0140}{\emph{PoS} {\bfseries CORFU2014}
  (2015) 140} [\href{https://arxiv.org/abs/1504.08114}{{\ttfamily
  1504.08114}}].

\bibitem{Howe:2015bdd}
P.S.~Howe and U.~Lindstr\"om, \emph{{Notes on super Killing tensors}},
  \href{https://doi.org/10.1007/JHEP03(2016)078}{\emph{JHEP} {\bfseries 03}
  (2016) 078} [\href{https://arxiv.org/abs/1511.04575}{{\ttfamily
  1511.04575}}].

\bibitem{Howe:2018lwu}
P.S.~Howe and U.~Lindstr\"om, \emph{{Some remarks on (super)-conformal
  Killing-Yano tensors}},
  \href{https://doi.org/10.1007/JHEP11(2018)049}{\emph{JHEP} {\bfseries 11}
  (2018) 049} [\href{https://arxiv.org/abs/1808.00583}{{\ttfamily
  1808.00583}}].

\bibitem{Kuzenko:2019tys}
S.M.~Kuzenko and E.S.N.~Raptakis, \emph{{Symmetries of supergravity backgrounds
  and supersymmetric field theory}},
  \href{https://doi.org/10.1007/JHEP04(2020)133}{\emph{JHEP} {\bfseries 04}
  (2020) 133} [\href{https://arxiv.org/abs/1912.08552}{{\ttfamily
  1912.08552}}].

\bibitem{Chandia:2022uyy}
O.~Chandia and B.C.~Vallilo, \emph{{Superspaces for heterotic pure spinor
  string compactifications}},
  \href{https://doi.org/10.1140/epjc/s10052-022-10947-0}{\emph{Eur. Phys. J. C}
  {\bfseries 82} (2022) 991}
  [\href{https://arxiv.org/abs/2205.01765}{{\ttfamily 2205.01765}}].

\bibitem{Figueroa-OFarrill:2007omz}
J.M.~Figueroa-O'Farrill, E.~Hackett-Jones and G.~Moutsopoulos, \emph{{The
  Killing superalgebra of ten-dimensional supergravity backgrounds}},
  \href{https://doi.org/10.1088/0264-9381/24/13/010}{\emph{Class. Quant. Grav.}
  {\bfseries 24} (2007) 3291}
  [\href{https://arxiv.org/abs/hep-th/0703192}{{\ttfamily hep-th/0703192}}].

\bibitem{Wald:1993nt}
R.M.~Wald, \emph{{Black hole entropy is the Noether charge}},
  \href{https://doi.org/10.1103/PhysRevD.48.R3427}{\emph{Phys. Rev. D}
  {\bfseries 48} (1993) R3427}
  [\href{https://arxiv.org/abs/gr-qc/9307038}{{\ttfamily gr-qc/9307038}}].

\bibitem{Jacobson:2015uqa}
T.~Jacobson and A.~Mohd, \emph{{Black hole entropy and Lorentz-diffeomorphism
  Noether charge}},
  \href{https://doi.org/10.1103/PhysRevD.92.124010}{\emph{Phys. Rev. D}
  {\bfseries 92} (2015) 124010}
  [\href{https://arxiv.org/abs/1507.01054}{{\ttfamily 1507.01054}}].

\bibitem{Bandos:2024pns}
I.~Bandos, P.~Meessen and T.~Ort\'\i{}n, \emph{{Noether-Wald and Komar charges
  in supergravity, fermions, and Killing supervectors in superspace}},
  \href{https://doi.org/10.1088/1742-6596/2912/1/012007}{\emph{J. Phys. Conf.
  Ser.} {\bfseries 2912} (2024) 012007}
  [\href{https://arxiv.org/abs/2411.01020}{{\ttfamily 2411.01020}}].

\bibitem{Gueven:1980be}
R.~Gueven, \emph{{Black holes have no superhair}},
  \href{https://doi.org/10.1103/PhysRevD.22.2327}{\emph{Phys. Rev. D}
  {\bfseries 22} (1980) 2327}.

\bibitem{Cordero:1978ud}
P.~Cordero and C.~Teitelboim, \emph{{Remarks on supersymmetric black holes}},
  \href{https://doi.org/10.1016/0370-2693(78)90352-0}{\emph{Phys. Lett. B}
  {\bfseries 78} (1978) 80}.

\bibitem{Gueven:1982tk}
R.~Gueven, \emph{{Extreme Reissner-Nordstrom black holes can support a
  'suprehair'}}, \href{https://doi.org/10.1103/PhysRevD.25.3117}{\emph{Phys.
  Rev. D} {\bfseries 25} (1982) 3117}.

\bibitem{Aichelburg:1983ux}
P.C.~Aichelburg and R.~Gueven, \emph{{Supersymmetric black holes in $N=2$
  supergravity theory}},
  \href{https://doi.org/10.1103/PhysRevLett.51.1613}{\emph{Phys. Rev. Lett.}
  {\bfseries 51} (1983) 1613}.

\bibitem{Liberati:2015xcp}
S.~Liberati and C.~Pacilio, \emph{{Smarr Formula for Lovelock Black Holes: a
  Lagrangian approach}},
  \href{https://doi.org/10.1103/PhysRevD.93.084044}{\emph{Phys. Rev. D}
  {\bfseries 93} (2016) 084044}
  [\href{https://arxiv.org/abs/1511.05446}{{\ttfamily 1511.05446}}].

\bibitem{Ortin:2021ade}
T.~Ort\'{\i}n, \emph{{Komar integrals for theories of higher order in the
  Riemann curvature and black-hole chemistry}},
  \href{https://doi.org/10.1007/JHEP08(2021)023}{\emph{JHEP} {\bfseries 08}
  (2021) 023} [\href{https://arxiv.org/abs/2104.10717}{{\ttfamily
  2104.10717}}].

\bibitem{Mitsios:2021zrn}
D.~Mitsios, T.~Ort\'\i{}n and D.~Pere\~n\'\i{}guez, \emph{{Komar integral and
  Smarr formula for axion-dilaton black holes versus S duality}},
  \href{https://doi.org/10.1007/JHEP08(2021)019}{\emph{JHEP} {\bfseries 08}
  (2021) 019} [\href{https://arxiv.org/abs/2106.07495}{{\ttfamily
  2106.07495}}].

\bibitem{Meessen:2022hcg}
P.~Meessen, D.~Mitsios and T.~Ort\'\i{}n, \emph{{Black hole chemistry, the
  cosmological constant and the embedding tensor}},
  \href{https://doi.org/10.1007/JHEP12(2022)155}{\emph{JHEP} {\bfseries 12}
  (2022) 155} [\href{https://arxiv.org/abs/2203.13588}{{\ttfamily
  2203.13588}}].

\bibitem{Pasti:1996vs}
P.~Pasti, D.P.~Sorokin and M.~Tonin, \emph{{On Lorentz invariant actions for
  chiral p forms}}, \href{https://doi.org/10.1103/PhysRevD.55.6292}{\emph{Phys.
  Rev. D} {\bfseries 55} (1997) 6292}
  [\href{https://arxiv.org/abs/hep-th/9611100}{{\ttfamily hep-th/9611100}}].

\bibitem{DallAgata:1997yxl}
G.~Dall'Agata, K.~Lechner and M.~Tonin, \emph{{Covariant actions for N=1, D = 6
  supergravity theories with chiral bosons}},
  \href{https://doi.org/10.1016/S0550-3213(97)00742-6}{\emph{Nucl. Phys. B}
  {\bfseries 512} (1998) 179}
  [\href{https://arxiv.org/abs/hep-th/9710127}{{\ttfamily hep-th/9710127}}].

\bibitem{DallAgata:1998ahf}
G.~Dall'Agata, K.~Lechner and M.~Tonin, \emph{{D = 10, N = IIB supergravity:
  Lorentz invariant actions and duality}},
  \href{https://doi.org/10.1088/1126-6708/1998/07/017}{\emph{JHEP} {\bfseries
  07} (1998) 017} [\href{https://arxiv.org/abs/hep-th/9806140}{{\ttfamily
  hep-th/9806140}}].

\bibitem{deWit:2005ub}
B.~de~Wit, H.~Samtleben and M.~Trigiante, \emph{{Magnetic charges in local
  field theory}},
  \href{https://doi.org/10.1088/1126-6708/2005/09/016}{\emph{JHEP} {\bfseries
  09} (2005) 016} [\href{https://arxiv.org/abs/hep-th/0507289}{{\ttfamily
  hep-th/0507289}}].

\bibitem{Ortin:2015hya}
T.~Ort\'{\i}n, \emph{{Gravity and Strings}}, Cambridge Monographs on
  Mathematical Physics, Cambridge University Press, 2nd ed.~ed. (7, 2015),
  \href{https://doi.org/10.1017/CBO9781139019750}{10.1017/CBO9781139019750}.

\bibitem{Bandos:2019lps}
I.~Bandos, \emph{{Superstring at the boundary of open supermembrane interacting
  with D=4 supergravity and matter supermultiplets}},
  \href{https://doi.org/10.1007/JHEP12(2019)106}{\emph{JHEP} {\bfseries 12}
  (2019) 106} [\href{https://arxiv.org/abs/1906.09872}{{\ttfamily
  1906.09872}}].

\bibitem{Neeman:1978njh}
Y.~Ne'eman and T.~Regge, \emph{{Gauge theory of gravity and supergravity on a
  group manifold}}, \href{https://doi.org/10.1007/BF02724472}{\emph{Riv. Nuovo
  Cim.} {\bfseries 1N5} (1978) 1}.

\bibitem{Castellani:1991eu}
L.~Castellani, R.~D'Auria and P.~Fr\'e, \emph{{Supergravity and superstrings: a
  geometric perspective. Vol. 2: Supergravity}}, World Scientific (1991).

\bibitem{Bandos:1995dw}
I.A.~Bandos, D.P.~Sorokin and D.~Volkov, \emph{{On the generalized action
  principle for superstrings and supermembranes}},
  \href{https://doi.org/10.1016/0370-2693(95)00506-G}{\emph{Phys. Lett. B}
  {\bfseries 352} (1995) 269}
  [\href{https://arxiv.org/abs/hep-th/9502141}{{\ttfamily hep-th/9502141}}].

\bibitem{Ballesteros:2023iqb}
R.~Ballesteros, C.~G\'omez-Fayr\'en, T.~Ort\'\i{}n and M.~Zatti, \emph{{On
  scalar charges and black hole thermodynamics}},
  \href{https://doi.org/10.1007/JHEP05(2023)158}{\emph{JHEP} {\bfseries 05}
  (2023) 158} [\href{https://arxiv.org/abs/2302.11630}{{\ttfamily
  2302.11630}}].

\bibitem{Cremmer:1980ru}
E.~Cremmer and S.~Ferrara, \emph{{Formulation of Eleven-Dimensional
  Supergravity in Superspace}},
  \href{https://doi.org/10.1016/0370-2693(80)90662-0}{\emph{Phys. Lett. B}
  {\bfseries 91} (1980) 61}.

\bibitem{Brink:1980az}
L.~Brink and P.S.~Howe, \emph{{Eleven-Dimensional Supergravity on the
  Mass-Shell in Superspace}},
  \href{https://doi.org/10.1016/0370-2693(80)91002-3}{\emph{Phys. Lett. B}
  {\bfseries 91} (1980) 384}.

\bibitem{Townsend:2001rg}
P.K.~Townsend and M.~Zamaklar, \emph{{The First law of black brane mechanics}},
  \href{https://doi.org/10.1088/0264-9381/18/23/320}{\emph{Class. Quant. Grav.}
  {\bfseries 18} (2001) 5269}
  [\href{https://arxiv.org/abs/hep-th/0107228}{{\ttfamily hep-th/0107228}}].

\bibitem{Haas:2014spa}
P.A.~Haas, \emph{{Smarr\textquoteright{}s formula in 11-dimensional
  supergravity}}, \href{https://doi.org/10.1088/1361-6382/aac44a}{\emph{Class.
  Quant. Grav.} {\bfseries 35} (2018) 135005}
  [\href{https://arxiv.org/abs/1405.3708}{{\ttfamily 1405.3708}}].

\bibitem{Dias:2019wof}
O.J.C.~Dias, G.S.~Hartnett and J.E.~Santos, \emph{{Covariant Noether charges
  for type IIB and 11-dimensional supergravities}},
  \href{https://doi.org/10.1088/1361-6382/abc136}{\emph{Class. Quant. Grav.}
  {\bfseries 38} (2021) 015003}
  [\href{https://arxiv.org/abs/1912.01030}{{\ttfamily 1912.01030}}].

\bibitem{Candiello:1993di}
A.~Candiello and K.~Lechner, \emph{{Duality in supergravity theories}},
  \href{https://doi.org/10.1016/0550-3213(94)90389-1}{\emph{Nucl. Phys. B}
  {\bfseries 412} (1994) 479}
  [\href{https://arxiv.org/abs/hep-th/9309143}{{\ttfamily hep-th/9309143}}].

\bibitem{DAuria:1982uck}
R.~D'Auria and P.~Fre, \emph{{Geometric Supergravity in d = 11 and Its Hidden
  Supergroup}}, \href{https://doi.org/10.1016/0550-3213(82)90281-4}{\emph{Nucl.
  Phys. B} {\bfseries 201} (1982) 101}.

\bibitem{Bandos:2004xw}
I.A.~Bandos, J.A.~de~Azcarraga, J.M.~Izquierdo, M.~Picon and O.~Varela,
  \emph{{On the underlying gauge group structure of D=11 supergravity}},
  \href{https://doi.org/10.1016/j.physletb.2004.06.079}{\emph{Phys. Lett. B}
  {\bfseries 596} (2004) 145}
  [\href{https://arxiv.org/abs/hep-th/0406020}{{\ttfamily hep-th/0406020}}].

\bibitem{Bandos:2004ym}
I.A.~Bandos, J.A.~de~Azcarraga, M.~Picon and O.~Varela, \emph{{On the
  formulation of D = 11 supergravity and the composite nature of its three-form
  gauge field}}, \href{https://doi.org/10.1016/j.aop.2004.11.016}{\emph{Annals
  Phys.} {\bfseries 317} (2005) 238}
  [\href{https://arxiv.org/abs/hep-th/0409100}{{\ttfamily hep-th/0409100}}].

\bibitem{Aichelburg:1984ef}
P.C.~Aichelburg and F.~Embacher, \emph{{Supercharge and background
  perturbations of multi-black hole systems}},
  \href{https://doi.org/10.1088/0264-9381/2/1/007}{\emph{Class. Quant. Grav.}
  {\bfseries 2} (1985) 65}.

\bibitem{Aneesh:2020fcr}
P.B.~Aneesh, S.~Chakraborty, S.J.~Hoque and A.~Virmani, \emph{{First law of
  black hole mechanics with fermions}},
  \href{https://doi.org/10.1088/1361-6382/aba5ab}{\emph{Class. Quant. Grav.}
  {\bfseries 37} (2020) 205014}
  [\href{https://arxiv.org/abs/2004.10215}{{\ttfamily 2004.10215}}].

\bibitem{Huebscher:2010ib}
M.~H{\"u}bscher, T.~Ort\'{\i}n and C.S.~Shahbazi, \emph{{The tensor hierarchies
  of pure N=2, d=4,5,6 supergravities}},
  \href{https://doi.org/10.1007/JHEP11(2010)130}{\emph{JHEP} {\bfseries 11}
  (2010) 130} [\href{https://arxiv.org/abs/1006.4457}{{\ttfamily 1006.4457}}].

\bibitem{Barnich:2001jy}
G.~Barnich and F.~Brandt, \emph{{Covariant theory of asymptotic symmetries,
  conservation laws and central charges}},
  \href{https://doi.org/10.1016/S0550-3213(02)00251-1}{\emph{Nucl. Phys. B}
  {\bfseries 633} (2002) 3}
  [\href{https://arxiv.org/abs/hep-th/0111246}{{\ttfamily hep-th/0111246}}].

\bibitem{Barnich:2003xg}
G.~Barnich, \emph{{Boundary charges in gauge theories: Using Stokes theorem in
  the bulk}}, \href{https://doi.org/10.1088/0264-9381/20/16/310}{\emph{Class.
  Quant. Grav.} {\bfseries 20} (2003) 3685}
  [\href{https://arxiv.org/abs/hep-th/0301039}{{\ttfamily hep-th/0301039}}].

\end{thebibliography}\endgroup

\end{document}